\documentclass[pra,superscriptaddress,twocolumn,amsmath,amssymb]{revtex4-1}

\usepackage[utf8]{inputenc}
\usepackage{times}

\usepackage{amsmath, amssymb, amsthm, mathtools}
\usepackage{physics, bm, bbold, upgreek, accents}
\usepackage{braket}
\usepackage{cancel}

\usepackage[pdftex]{graphicx}
\DeclareGraphicsExtensions{.png, .pdf, .eps, .jpg}
\graphicspath{{./figs/}}
\usepackage{subfigure}
\usepackage{tabularx}
\usepackage{array, longtable, multirow, tabu, hhline}
\newcolumntype{M}[1]{>{\centering\arraybackslash}m{#1}}
\newcolumntype{N}{@{}m{0pt}@{}}

\usepackage{xcolor, color, soul}
\usepackage[normalem]{ulem}

\usepackage{tikz}
\usetikzlibrary{
	decorations.pathmorphing,
	calc,
	shapes.geometric,
	positioning,
	arrows.meta
}
\tikzstyle{box}   = [draw, fill=blue!10, text centered, rectangle, minimum width=2cm, minimum height=1cm]
\tikzstyle{arrow} = [thick, ->, >=stealth]

\usepackage{tcolorbox}
\usepackage{placeins}
\usepackage{xifthen} 

\usepackage{orcidlink}
\usepackage{hyperref}
\definecolor{myurlcolor}{rgb}{0, 0, 0.7}
\definecolor{myrefcolor}{rgb}{0.8, 0, 0}
\hypersetup{
	unicode=true,
	pdfusetitle,
	bookmarks=false,
	breaklinks=false,
	pdfborder={0 0 0},
	backref=false,
	colorlinks=true,
	linkcolor=myrefcolor,
	citecolor=myurlcolor,
	urlcolor=myurlcolor
}

\theoremstyle{plain}

\ifx\proof\undefined
\newenvironment{proof}[1][\proofname]{\par
	\normalfont\topsep6\p@\@plus6\p@\relax
	\trivlist
	\itemindent\parindent
	\item[\hskip\labelsep\scshape #1]\ignorespaces
}{
	\endtrivlist\@endpefalse
}
\providecommand{\proofname}{Proof}
\fi

\newcommand{\eqnref}[1]{Eq.~(\ref{#1})}

\newcommand{\figref}[1]{Fig.~\ref{#1}}
\newcommand{\tabref}[1]{Table~\ref{#1}}
\newcommand{\secref}[1]{Sec.~\ref{#1}}
\newcommand{\appref}[1]{App.~\ref{#1}}

\newcommand{\smat}[1]{{\mathsf{#1}}}

\newcommand{\avg}[1]{\langle #1 \rangle}

\newcommand{\qsvec}[1]{{\hat{\smat{#1}}}}
\newcommand{\Id}{\smat{I}}	 

\newcommand{\tp}{{\rm T}}

\renewcommand\Re{\operatorname{Re}}
\renewcommand\Im{\operatorname{Im}}
\renewcommand\Tr{\operatorname{Tr}}

\newcommand{\mrm}[1]{\mathrm{#1}}

\newcommand{\bpm}{\begin{pmatrix}}
	\newcommand{\epm}{\end{pmatrix}}
\newcommand{\beq}{\begin{equation}}
	\newcommand{\eeq}{\end{equation}}
\newcommand{\ba}{\begin{align}}
	\newcommand{\ea}{\end{align}}
\newcommand{\bi}{\begin{itemize}}
	\newcommand{\ei}{\end{itemize}}

\newcommand{\bvec}[1]{\hat{\bm{#1}}}

\renewcommand{\H}{\textbf{H}}
\newcommand{\K}{\textbf{K}}

\newcommand{\q}{\bvec{q}}
\newcommand{\p}{\bvec{p}}

\newcommand{\n}{\textbf{n}}

\renewcommand{\i}{\mrm{i}}

\newcommand{\gzero}{g_{\scriptscriptstyle 0}}
\newcommand{\dzero}{\Delta_{\scriptscriptstyle 0}}
\begin{document}
	\title{Optomechanically controlled response amplification for enhanced quantum sensing}

\author{Javid Naikoo \orcidlink{0000-0001-7552-7200}}
	\email{javid.naikoo@upol.cz}
	\affiliation{Joint Laboratory of Optics of Palacký University and Institute of Physics ASCR, Faculty of Science, Palacký University,
		Czech Republic, 17. listopadu 12, 779 00, Olomouc, Czech Republic}

\begin{abstract}
	We show that strongly amplified dynamical responses in cavity optomechanical systems can be harnessed for enhanced quantum sensing. By tuning the optomechanical interaction to a regime of enhanced susceptibility, weak perturbations produce disproportionately large changes in the system response, leading to substantially improved estimation precision. Using Gaussian  estimation theory, we demonstrate that the quantum Fisher information exhibits a divergent scaling as the perturbation strength decreases, implying a corresponding suppression of the estimation error. 
	We further show that heterodyne detection of the output cavity field yields the classical Fisher information with the same asymptotic scaling as the quantum Fisher information, demonstrating that the enhanced sensitivity is accessible with a standard measurement protocol. These results identify amplified optomechanical dynamics as a controllable resource for  quantum enhanced  sensing and metrology.
\end{abstract}

	\maketitle

\section{Introduction}
The pursuit of increasingly precise measurements has long been a driving force in the development of physics, enabling stringent tests of fundamental principles and the detection of exceedingly weak physical effects. Quantum sensing has emerged as a prominent application of quantum technologies, exploiting resources such as superposition and entanglement, and other nonclassical effects to achieve measurement sensitivities that surpass classical limits~\cite{Giovannetti2004,Degen2017,Demkowicz2012,Chaves2013}.  At the same time, there is growing interest in understanding how quantum mechanics manifests at larger scales, as demonstrated by experiments showing coherence in systems ranging from superconducting circuits to matter-wave interference with complex molecules~\cite{Andrews1997,Friedman2000,Eibenberger2013,Fein2019}. These developments have motivated the exploration of mesoscopic platforms that not only bridge the quantum-to-classical transition but also serve as functional devices for precision measurement and quantum-enhanced technologies.

Among such platforms, cavity optomechanical systems have emerged as a leading candidate for studying macroscopic quantum behavior~\cite{Aspelmeyer2014}. In these systems, a mechanical resonator couples to an optical cavity field  via radiation pressure. The mechanical displacement shifts the cavity resonance frequency, while the intracavity  field exerts a back-action force on the mechanical element. This mutual interaction enables highly sensitive optical readout of mechanical motion as well as precise optical control of  mechanical dynamics~\cite{Verhagen2012,Chen2013,Aspelmeyer2014,Bowen2015}. Continuous experimental progress has enabled cooling of  mechanical resonators close to their quantum ground state, thereby accessing regimes where \emph{quantum fluctuations} play a central role in their dynamics~\cite{Chan2011,Teufel2011,Peterson2016}.

Beyond their fundamental interest, optomechanical systems constitute powerful platforms for precision force and displacement sensing~\cite{Aspelmeyer2014}. Their high susceptibility to external perturbations enables the detection of extremely weak forces, with applications ranging from precision inertial sensing and gravimetry to searches for faint signals such as ultralight dark matter~\cite{Mow2008,Abbott2009,Matsumoto2019,Pierce2018}. Significant efforts have therefore been devoted to improving their sensing capabilities through advanced measurement and control techniques. Recent studies have demonstrated enhanced performance using photon-counting-based inference protocols~\cite{LewisClark2022}, Kerr-assisted optomechanical architectures~\cite{Cheng2025}, and entanglement-enhanced sensing schemes~\cite{Li2026}.

Of particular relevance here, optomechanical systems exhibit nonlinear dynamical regimes associated with sharp transitions between distinct states. Owing to the radiation-pressure interaction, these systems can exhibit bistability, self-sustained oscillations, limit cycles, and chaotic dynamics~\cite{Ludwig2008,Niels2014,Bakermeier2015,Zhu2023}. Such phenomena are often accompanied by transitions between distinct dynamical regimes, where the response of the system becomes exceptionally sensitive to small parameter variations. The resulting enhancement of susceptibility near transition points is closely related to critical phenomena in nonequilibrium systems and provides a natural mechanism for amplifying weak perturbations~\cite{Wang2016,Simon2019,Chen2021,Xiong2023}. Motivated by these observations, criticality-enhanced sensing in optomechanical platforms has recently been proposed~\cite{Tang2023}. By operating in the vicinity of dynamical critical points, such schemes harness the amplified response associated with critical behavior to facilitate the detection of weak perturbations.

In this work, we investigate quantum-enhanced sensing in a lossy cavity optomechanical system, accounting for dissipation arising from both intrinsic system–environment coupling and the probing channels. To characterize the achievable precision, we consider Gaussian probe states and employ Gaussian  estimation theory, with the Fisher information serving as the central figure of merit. This approach allows us to relate the ultimate estimation precision to the Green’s function of the driven optomechanical system. We show that the optomechanical coupling can induce pronounced response amplification associated with singular behavior of the Green's function near dynamical transitions, leading to enhanced scaling of the achievable precision.

The paper is organized as follows. In Sec.~\ref{sec:model}, we introduce the cavity optomechanical system and develop its linearized dynamical description, along with the corresponding input--output framework for the relevant observables. In Sec.~\ref{sec:sensitivity}, we analyze the sensitivity of the system to external perturbations, formulating the estimation problem in terms of the Green’s function and associated statistical quantities. Section~\ref{sec:optoSensing} then demonstrates how optomechanical coupling can be used to engineer singular response, leading to enhanced sensing performance and nontrivial scaling of estimation precision. We conclude in Sec.~\ref{sec:concl}.

\section{Cavity Optomechanical Model and Linearized Dynamics}\label{sec:model}
\begin{figure}[t]
	\includegraphics[width=0.95\linewidth]{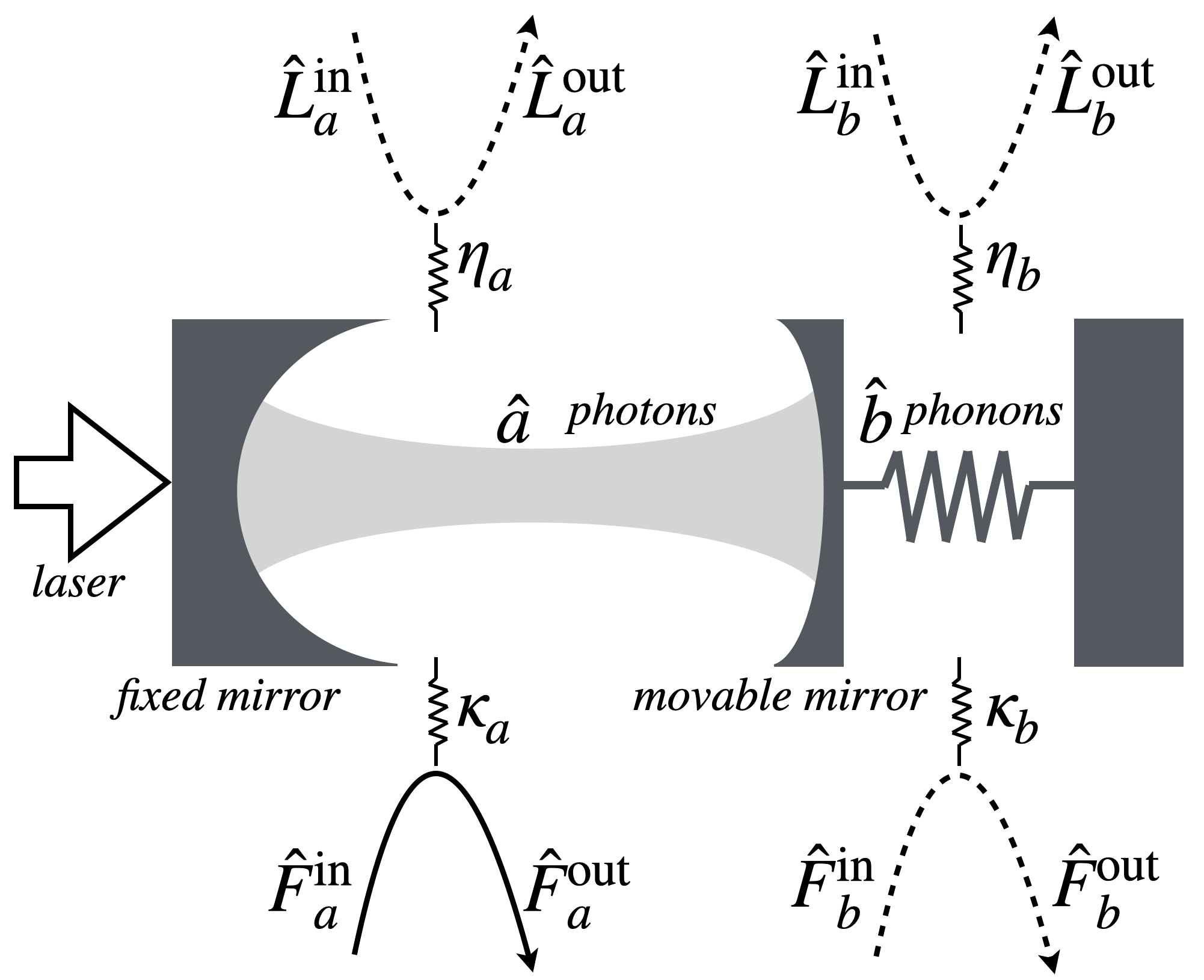}
\caption{\textbf{Schematic of the cavity optomechanical system}. The setup consists of an optical cavity mode $\hat{a}$ coupled via radiation pressure to a mechanical (phonon) mode $\hat{b}$ associated with the movable cavity element. The  optical probe channel $\hat{F}_{a}^{\mrm{in(out)}}$, coupled with strength $\kappa_a$, represents the experimentally  \emph{accessible} input (output) field used for parameter estimation. The remaining channels, namely $\hat{F}_{b}^{\mrm{in(out)}}$ and the bath fields $\hat{L}_{a,b}^{\mrm{in(out)}}$, with coupling rates $\kappa_b$ and $\eta_{a,b}$, are fictitious/uncontrollable \emph{inaccessible} channels introduced within the effective input--output description of the system.}
	\label{fig:model}
\end{figure}

We consider a cavity optomechanical system consisting of a single optical mode coupled to a mechanical oscillator, as schematically shown in~\figref{fig:model}. The system is described by the Hamiltonian
\begin{align}
	H &=  \omega_\mrm{cav}\, \hat{a}^{\dagger} \hat{a}	+ \omega_\mrm{M}\, \hat{b}^{\dagger} \hat{b}
	- g_{0}\, \hat{a}^{\dagger} \hat{a} (\hat{b}^{\dagger} + \hat{b})
	\nonumber \\& \quad + \i \, \varepsilon_\mrm{L}\, (e^{-\i\, \omega_\mrm{L}} \hat{a}^{\dagger}
	- e^{\i \, \omega_\mrm{L}} \hat{a}),
\end{align}
where $\hat{a}$ ($\hat{a}^{\dagger}$) and $\hat{b}$ ($\hat{b}^{\dagger}$) denote the annihilation (creation) operators for the optical cavity mode and the mechanical oscillator, respectively. The parameters $\omega_\mrm{cav}$ and $\omega_\mrm{M}$ represent the bare resonance frequencies of the optical and mechanical modes. The term proportional to $\gzero$ describes the radiation-pressure interaction, whereby the photon number $\hat{a}^{\dagger}\hat{a}$ couples to the mechanical position quadrature $(\hat{b}^{\dagger}+\hat{b})$. The final term corresponds to coherent driving of the cavity by an external laser field of amplitude $\epsilon_\mrm{L}$ and frequency $\omega_\mrm{L}$.

Before proceeding further, it is convenient to separate the coherent dynamics from the quantum fluctuations. We decompose the cavity and mechanical operators as
\begin{equation}
	\delta \hat{a} \equiv \hat{a} - \alpha, \quad 	\delta \hat{b} \equiv \hat{b} - \beta, \label{eq:alphabeta}
\end{equation}
where $\alpha\!=\! \langle \hat{a} \rangle$ and $\beta\!=\! \langle \hat{b} \rangle$ denote the corresponding expectation values. The operators $\delta \hat{a}$ and $\delta \hat{b}$ thus describe fluctuations about the mean fields and capture the quantum noise of the system. In the regime of strong driving, the dynamics is dominated by the mean fields $\alpha$ and $\beta$, while the fluctuations remain small. This permits a systematic expansion in the fluctuation operators, retaining only linear terms and yielding an effective linearized description of the coupled optical and mechanical modes.

For notational simplicity, we omit the $\delta$ symbol henceforth and relabel the fluctuation operators $\delta \hat{a}$ and $\delta \hat{b}$ as $\hat{a}$ and $\hat{b}$; it is understood throughout that these operators represent deviations from their respective mean values.

\subsection{Quantum Langevin equations}
In practice, both optical and mechanical modes interact with environmental reservoirs, leading to photon loss from the cavity and phonon damping in the mechanical oscillator. Incorporating these dissipative processes leads to the quantum Langevin equations governing the dynamics of the fluctuation operators~\cite{GardinerBook}. These equations read (see ~\appref{app:model} for details)
\begin{align}
	\partial_t \hat{a} &\!=\!
	\left(\i \, \Delta -\gamma_{a} \right)\hat{a}
	+\i \, g(\hat{b}^{\dagger}+\hat{b})
	+\sqrt{\kappa_a}\hat{F}_{a}^\mrm{in} +\sqrt{\eta_a} \hat{L}_{a}^\mrm{in},
	\label{eq:a}
	\\
	\partial_t \hat{b} &\!=\!
	\left(-\i\, \omega_b - \gamma_{b} \right)\hat{b}
	+\i \, g (\hat{a}^{\dagger}+\hat{a})
	+\sqrt{\kappa_b}\hat{F}_{b}^\mrm{in} +\sqrt{\eta_b}\hat{L}_{b}^\mrm{in},
	\label{eq:b}
\end{align}
Here, $\Delta \!=\! \dzero - 2 \gzero  \Re(\beta)$ denotes the effective cavity detuning, where $\dzero \!=\! \omega_\mrm{L} - \omega_\mrm{cav}$ is the bare detuning between the driving laser frequency and the cavity resonance frequency. The quantity $g \!=\! \gzero |\alpha|$ represents the linearized optomechanical coupling strength, with $\alpha$ and $\beta$ introduced in~\eqnref{eq:alphabeta}.

The system couples to external probe fields and internal reservoirs through two distinct channels. The external channels, characterized by the rates $\kappa_{\ell}$, inject probe fields into the system via the input operators $\hat{F}_{\ell}^{\mathrm{in}}$, while the intrinsic dissipation channels, with rates $\eta_{\ell}$, give rise to environmental noise described by $\hat{L}_{\ell}^{\mathrm{in}}$. These processes jointly account for both driven fluctuations and irreversible coupling to internal baths. The combined effect of these channels determines the total damping of each mode. In particular, for $\ell = a,b$, the overall damping rate is given by $\gamma_{\ell} = (\kappa_{\ell} + \eta_{\ell})/2$. Consequently, the system dynamics incorporates both coherent optomechanical interactions and dissipative fluctuations arising from external driving and intrinsic losses.

To streamline the notation and make the structure of the coupled equations more transparent, it is convenient to collect the operators into column vectors. We therefore define
\begin{align}
	\bvec{a}&\!:=\!
	\begin{pmatrix}
		\hat a\vspace{1mm} \\
		\hat b
	\end{pmatrix}, \qquad ~
	\bvec{a}^{\dagger} \!:=\!
	\begin{pmatrix}
		\hat a^{\dagger}\vspace{1mm} \\
		\hat b^{\dagger}
	\end{pmatrix} \ne (\hat{a}^\dagger ~~\hat{b}^\dagger),
	\nonumber\\
	\bvec{F}^\mrm{in}&\!:=\!
	\begin{pmatrix}
		\hat{F}_{a}^\mrm{in}\vspace{1mm} \\
		\hat{F}_{b}^\mrm{in}
	\end{pmatrix}, \quad
	\bvec{L}^\mrm{in} \!:=\!
	\begin{pmatrix}
		\hat{L}_{a}^\mrm{in}\vspace{1mm} \\
		\hat{L}_{b}^\mrm{in}
	\end{pmatrix}.
\end{align}
Here $\bvec{a}$ groups together the fluctuation operators of the optical and mechanical modes, while $\bvec{a}^\dagger$ denotes the corresponding vector of creation operators written in the same column form. The distinction from the row vector $(\hat{a}^\dagger~~\hat{b}^\dagger)$ will be important when writing the equations in matrix form. Similarly, $\bvec{F}^\mrm{in}$ and $\bvec{L}^\mrm{in}$ collect the probe and bath input fields, respectively.  In terms of these vectors, the coupled quantum Langevin Eqs.~\eqref{eq:a} and \eqref{eq:b} can be written in a compact matrix form,
\begin{equation}\label{eq:veca}
	\partial_t \bvec{a}
	\!=\!
	-\i\,\H \, \bvec{a}
	+\K_\mrm{om} \, \bvec{a}^{\dagger}
	+\K_\mrm{probe}\,  \bvec{F}^\mrm{in}
	+\K_\mrm{bath} \, \bvec{L}^\mrm{in},
\end{equation}
which makes explicit the different contributions to the dynamics. The first term describes the coherent evolution governed by an effective non-Hermitian dynamical matrix $\H$, the second term captures the parametric coupling between operators and their Hermitian conjugates arising from the linearized optomechanical interaction, while the remaining terms account for the  probe and environmental noise. The matrices appearing in Eq.~\eqref{eq:veca} are given by
\begin{align}\label{eq:MatricesTimeDomain}
	\H&=
	\begin{pmatrix}
		-\Delta -\i \, \gamma_{a} & -g\\
		-g& \omega_b -\i \, \gamma_{b}
	\end{pmatrix}, \quad 	\K_{\rm om} = \begin{pmatrix}
		0 & \i \, g\\
		\i \, g& 0
	\end{pmatrix},
	\nonumber\\
	\K_{\rm probe}&=
	\begin{pmatrix}
		\sqrt{\kappa_a} & 0\\
		0 & \sqrt{\kappa_b}
	\end{pmatrix},
	\quad
	\K_{\rm bath}=
	\begin{pmatrix}
		\sqrt{\eta_a} & 0\\
		0 & \sqrt{\eta_b}
	\end{pmatrix}.
\end{align}
Here, $\H$ encodes both the coherent detuned dynamics and the damping of the two modes, $\K_{\rm om}$ represents the linearized optomechanical interaction, and the matrices $\K_{\rm probe}$ and $\K_{\rm bath}$ specify the coupling strengths to the probe and loss channels, respectively.  In this form, the equations provide a convenient starting point for analyzing the system response in both time and frequency domains.

\subsection{Quadrature representation and output field statistics}
In order to analyze and quantify the measurement sensitivity, it is convenient to reformulate the dynamics in terms of field quadratures. This expresses the bosonic modes in terms of canonical Hermitian operators corresponding to field amplitudes and provides a natural framework for treating Gaussian states and their noise properties.

In the Fourier domain we define the generalized position and momentum quadratures associated with the field operator $\bvec a[\omega]$ as \footnote{For a vector of bosonic operators $\hat{\bvec{a}}(t)\!=\!\big(\hat{a}(t),\hat{b}(t),\ldots\big)^\tp$, we define the Fourier transform as $\hat{\bvec{a}}[\omega]\!:=\!\mathcal{F}_{\omega}[\hat{\bvec{a}}]\!=\!\int dt\,e^{\i\omega t}\hat{\bvec{a}}(t)$. The corresponding quadratures in frequency space are $\q[\omega]\!=\!\hat{\bvec{a}}[\omega]+(\hat{\bvec{a}}[\omega])^{\dagger}$ and $\p[\omega]\!=\!-\i\big(\hat{\bvec{a}}[\omega]-(\hat{\bvec{a}}[\omega])^{\dagger} \big)$, where the Hermitian conjugation acts entry-wise. Importantly, one has $\big(\hat{\bvec{a}}[\omega]\big)^{\dagger}\!=\!\hat{\bvec{a}}^{\dagger}[-\omega]$.} 
\begin{equation}\label{eq:quad_def}
	\q[\omega]\!=\!\bvec a[\omega]+(\bvec a[\omega])^{\dagger},\quad 
	\p[\omega]\!=\!-\i\bigl(\bvec a[\omega]-(\bvec a[\omega])^{\dagger}\bigr).
\end{equation}
These operators are Hermitian and correspond to the amplitude and phase quadratures of the field modes. In this representation the fluctuations of the optical field and their correlations can be conveniently expressed in terms of covariance matrices.

We now rewrite the dynamical equations in terms of these quadratures. Taking the Fourier transform of Eq.~(\ref{eq:veca}) together with its Hermitian conjugate and using the definitions in Eq.~(\ref{eq:quad_def}), we obtain a set of coupled equations for the quadrature components (see \appref{app:QuadRep}). It is convenient to express these equations using the canonical symplectic structure, encoded in the matrix $\smat{\Omega} \!=\! [\textbf{0}, \textbf{I}; -\textbf{I}, \textbf{0}]$, with $\textbf{0}$ and $\textbf{I}$ denoting the $2\times 2$ null and identity matrices, respectively. This matrix plays the role of the imaginary unit in  quadrature space, satisfying $\smat{\Omega}^2 \!=\! \smat{I}$~\cite{Ferraro2005}. In matrix form, the equations read
\begin{align}
	\omega\, \smat{\Omega}
	\begin{pmatrix}
		\q[\omega] \vspace{1mm} \\
		\p[\omega]
	\end{pmatrix}
	&=
	\smat{H} \,\smat{\Omega}
	\begin{pmatrix}
		\q[\omega] \vspace{1mm} \\
		\p[\omega]
	\end{pmatrix}
	+\smat K_{\rm om}
	\begin{pmatrix}
		\q[-\omega] \vspace{1mm} \\
		\p[-\omega]
	\end{pmatrix}
	\nonumber\\
	&
	+\smat K_{\rm probe}
	\begin{pmatrix}
		\q_\mrm{F}^\mrm{in}[\omega]\vspace{1mm} \\
		\p_\mrm{F}^\mrm{in}[\omega]
	\end{pmatrix}
	+\smat K_{\rm bath}
	\begin{pmatrix}
		\q_\mrm{L}^\mrm{in}[\omega] \vspace{1mm} \\
		\p_\mrm{L}^\mrm{in}[\omega]
	\end{pmatrix}.
	\label{eq:QuadPlusOmega}
\end{align}
The matrices appearing above admit the following block-structured representation
\begin{align}\label{eq:HKsFourierSpace}
	\smat{H} \!&=\! \mathcal{R}[\H],\quad 
	\smat{K}_{\rm om} \!=\! \mathcal{T}[\K_{\rm om}], \quad 
	\smat{K}_{\bullet} \!=\! \mathrm{diag}(\K_{\bullet},\K_{\bullet}),
\end{align}
where $\bullet$ labels the external channels, corresponding to the probe and bath contributions. The mappings $\mathcal{R}[\cdot]$ and $\mathcal{T}[\cdot]$ are defined as
\begin{align}
	\mathcal{R}[X] &\!:=\!
	\begin{pmatrix}
		\Re(X) & -\Im(X) \vspace{1mm} \\
		\Im(X) & \quad \Re(X)
	\end{pmatrix}, \label{eq:Rmap}\\ 
	\mathcal{T}[X] &\!:=\!
	\begin{pmatrix}
		\Re(X) & \quad \Im(X) \vspace{1mm} \\
		\Im(X) & -\Re(X)
	\end{pmatrix}. \label{eq:Tmap}
\end{align}
%
%
The first term on the right-hand side of \eqnref{eq:QuadPlusOmega} describes the intrinsic dynamics of the system, while the second term couples the positive and negative frequency components due to the optomechanical interaction. The remaining terms account for the input fluctuations originating from the probe and the inaccessible channels.

For notational convenience, we assemble all system and input quadratures into vector form,
\begin{equation}\label{eq:FSnotation}
	\qsvec S[\pm \omega] \!:=\!
	\begin{pmatrix}
		\q[\pm \omega]\vspace{1mm} \\
		\p[\pm \omega]
	\end{pmatrix}, ~
	\qsvec{S}^\mrm{in/out}_{\ast}[ \omega] \!:=\! \begin{pmatrix}
		\q_{\ast}^\mrm{in/out}[\omega] \vspace{1mm} \vspace{1mm} \\
		\p_{\ast}^\mrm{in/out}[\omega]
	\end{pmatrix}, ~ \ast = \mrm{F/L},
\end{equation}
where F/L pertains to the probe/inaccessible channel. In terms of these variables, the linear system in Eq.~(\ref{eq:QuadPlusOmega}) can be formally solved in the frequency domain. The resulting solution can be expressed as
\begin{align}\label{eq:Sw2}
	\qsvec{S}[\omega] &\!=\! \smat{G}[\omega] \, \smat{K}_\mrm{probe} \,\qsvec{S}_\mrm{F}^\mrm{in}[\omega] 
	+ \smat{G}[\omega] \, \smat{K}_\mrm{bath} \,\qsvec{S}_\mrm{L}^\mrm{in}[\omega] 
	\nonumber \\& \quad - \smat{G}[\omega] \, \smat{T}_\mrm{op}\,\qsvec{S}_\mrm{F}^\mrm{in}[-\omega] - \smat{G}[\omega] \, \smat{T}_\mrm{ob}\,\qsvec{S}_\mrm{L}^\mrm{in}[-\omega],
\end{align}
where the optomechanical-probe  and optomechanical-bath  transfer  matrices are defined as 
\begin{align}
 \smat{T}_\mrm{op} &\!:=\! 	\smat{K}_\mrm{om}\,\Omega^{-1}(\omega \Id+\smat{H})^{-1}
	\smat{K}_\mrm{probe}, \label{eq:Top}\\
	\smat{T}_\mrm{ob} &\!:=\! 	\smat{K}_\mrm{om}\,\Omega^{-1}(\omega \Id+\smat{H})^{-1} 	\smat{K}_\mrm{bath}, \label{eq:Tob}
\end{align}	
	 and 
\begin{equation}\label{eq:Gfun}
	\smat{G}[\omega]\!:=\! 
	\left[
	(\omega\, \Id-\smat{H})\,\Omega
	-\smat{K}_{\rm om}\, \Omega\,
	(\omega\, \Id+\smat{H})^{-1}
	\smat{K}_{\rm om}
	\right]^{-1},
\end{equation}
acts as the frequency-domain Green's function of the system. It fully characterizes the linear response of the quadratures to external probe and noise sources. Here, $\smat{I}$ denotes the $4\times 4$ identity matrix.  

To connect the internal dynamics with experimentally measurable quantities, we now relate the intracavity quadratures to the outgoing probe field through the standard input--output formalism~\cite{Gardiner1985},
\begin{equation}\label{eq:IOrel}
	\qsvec{S}_\mrm{F}^\mrm{out}[\omega]
	=
	\qsvec{S}_\mrm{F}^\mrm{in}[\omega]
	-
	\smat{K}_\mrm{probe}\,\qsvec{S}[\omega].
\end{equation}
Substituting the intracavity solution given in \eqnref{eq:Sw2} into \eqnref{eq:IOrel}  expresses the output field entirely in terms of the system Green's function 
\begin{align}
	 \qsvec{S}_\mrm{F}^\mrm{out}[\omega]  \!&=\! \left(\Id   - \kappa\, \smat{G}[\omega]\right) \, \qsvec{S}_\mrm{F}^\mrm{in}[\omega]  - \sqrt{\kappa} \, \smat{G}[\omega] \, \smat{K}_\mrm{bath} \,\qsvec{S}_\mrm{L}^\mrm{in}[\omega] \nonumber \\& \quad 
	-\sqrt{\kappa} \, \smat{G}[\omega] \, \smat{T}_\mrm{op} \,\qsvec{S}_\mrm{F}^\mrm{in}[-\omega] -\sqrt{\kappa} \, \smat{G}[\omega] \, \smat{T}_\mrm{ob} \,\qsvec{S}_\mrm{L}^\mrm{in}[-\omega],\label{eq:SFout}
\end{align}
thereby making explicit how the internal response is transferred to observable probe signals. Here we have assumed  $\kappa \!:=\! \kappa_a \!=\! \kappa_b$  the equal coupling strength of the probe fields to the photon and phonon  cavity modes, respectively, such that $\smat{K}_\mrm{probe} = \kappa \, \Id$ i.e. the photon and phonon field will suffer equal loss due to interaction with the probe fields.

We next turn to a statistical description of the output field in terms of its first and second moments. Denoting amplitude expectation values by $\smat{S}\!:=\!\langle\qsvec{S}\rangle$ and the symmetric covariance matrix $\smat{V}\!:=\! \langle  \tfrac{1}{2}\{\qsvec{S}_j, \qsvec{S}_k\} \rangle  - \langle \qsvec{S}_j\rangle \langle \qsvec{S}_k \rangle$. Since the quadrature vectors introduced above describe fluctuations, all first moments vanish; in particular
\begin{equation}\label{eq:Svanish}
\smat{S}_\mrm{F}^\mrm{out}[\omega] = 0.
\end{equation}
 The fluctuation properties are therefore encoded entirely in the covariance matrix of the output quadratures and is given by 
\begin{align}
	\smat{V}_\mrm{F}^\mrm{out}[\omega] &\!=\!
	(\Id-\kappa\, \smat{G}[\omega])
	\,\smat{V}_\mrm{F_{+}}^\mrm{in}
	(\Id-\kappa \, \smat{G}[\omega])^\tp
	\nonumber \\& \quad +
	\kappa \, \smat{G}[\omega]\,\Xi
	\,\smat V_{\mrm{L_{\pm}, F_{-}}}^\mrm{in}  \Xi^\tp
	\smat{G}[\omega]^\tp. \label{eq:VFout}
\end{align}
Here, $\smat{V}_{\mrm{F_{\pm}}}^\mrm{in} \!=\! \smat{V}_\mrm{F}^\mrm{in}[\pm\omega]$ and  $\smat{V}_{\mrm{L{\pm}, F_{-}}}^\mrm{in} \!=\! \smat{V}_\mrm{L}^\mrm{in}[\omega] \oplus \smat{V}_\mrm{F}^\mrm{in}[-\omega] \oplus \smat{V}_\mrm{L}^\mrm{in}[-\omega]$
with the first term in \eqnref{eq:VFout} describing the transformation of probe fluctuations by the system response, while the second term accounts for additional noise entering through the bath and the negative-frequency sector via the $4 \times 12$  optomechanical coupling matrix $\Xi := (\smat{K}_\mrm{bath} \vert \smat{T}_\mrm{op} \vert \smat{T}_\mrm{ob})$.

With these observable output quantities established, we are now in position to examine how an external perturbation modifies them through its effect on the Green's function, as developed in the following section.

\section{Sensitivity analysis}\label{sec:sensitivity}

\subsection{Perturbation of system parameters}
We now consider the system’s response to a weak external perturbation characterized by a small parameter $\theta$. The perturbation is implemented directly at the level of the dynamical matrix $\H$ given in \eqnref{eq:MatricesTimeDomain} as follows
\begin{equation}\label{eq:PerturbTimeDomain}
	\H \rightarrow \H - \theta\, \n,
\end{equation}
where $\n$ encodes the structure of the perturbation across the system degrees of freedom, specifying which entries (and therefore which physical parameters of $\H$) are affected. In this sense, $\n$ acts as a selection or weighting operator that identifies the subspace, modes, or coupling channels to which the perturbation is applied, thereby determining how the external influence is distributed within the system.

In quadrature space, we note that the counterpart $\smat{H}$ of $\H$ (see \eqnref{eq:HKsFourierSpace}) is modified as
\begin{equation}\label{eq:PerturbFourierDomain}
	\smat{H} \;\rightarrow\; \smat{H} -  \theta\, \smat{n},
\end{equation}
where $\smat{n}$ denotes the real representation of the perturbation matrix in the doubled Fourier-space basis, explicitly given by
\begin{equation}
	\smat{n}
	= \mathcal{R}[\n],
\end{equation}
with $\mathcal{R}$  defined in \eqnref{eq:Rmap}. This construction preserves the action of the original complex perturbation matrix $\n$ when expressed in the real-valued block form adopted in Fourier space.

As a consequence of this shift in the dynamical matrix, the Green's function introduced in \eqnref{eq:Gfun} is modified into the perturbed resolvent
\begin{equation}\label{eq:Gfun1}
	\smat{G}_{\theta}
	= -\,\Omega \Big[
	\smat{H}_{-} + \theta \, \smat{n}
	+ \smat{K}_{\rm om}' (\theta \, \smat{n} - \smat{H}_{+})^{-1} \smat{K}_{\rm om}'
	\Big]^{-1},
\end{equation}
where we have dropped the explicit $\omega$ dependence and  introduced the shorthand notations
\begin{equation}\label{eq:Hpm}
	\smat{H}_{\pm} := \omega\, \Id \pm \smat{H}, \quad \smat{K}_{\rm om}' := \smat{K}_{\rm om}\, \Omega.
\end{equation}
The \eqnref{eq:Gfun1} explicitly captures how the external perturbation alters the system response in frequency space through both the direct modification of $\smat{H}$ and its indirect effect on the coupling-mediated term.
At this stage, the structure of the problem becomes particularly transparent: the perturbation enters not only directly through the term $\theta\,\smat{n}$, but also indirectly through the inverse resolvent in the coupled sector.

To proceed analytically, we assume that $\smat{H}_{+}$ is invertible and that $\theta$ is sufficiently small to justify a perturbative expansion. We therefore expand the resolvent using a Neumann series. To first order in $\theta$, one finds
\begin{equation}
	(\theta\,\smat{n} - \smat{H}_{+})^{-1}
	= -\smat{H}_{+}^{-1}
	- \theta\,\smat{H}_{+}^{-1}\smat{n}\smat{H}_{+}^{-1}
	+ \mathcal{O}(\theta^2).
\end{equation}
Substituting this expansion back into \eqnref{eq:Gfun1}  yields
\begin{align}\label{eq:Gtheta}
	\smat{G}_{\theta}
	\approx & -\Omega \left[ \smat{H}_\mrm{eff} + \theta\, \smat{n}_\mrm{eff} \right]^{-1},
\end{align}
where the effective dynamical generator and the perturbation matrix are respectively given as
\begin{align}
	 \smat{H}_\mrm{eff} &\!:=\! \smat{H}_{-}  +  \smat{K}_{\rm om}' \smat{H}_{+}^{-1} \smat{K}_{\rm om}', \label{eq:Heff} \\
	  \smat{n}_\mrm{eff} &\!:=\!  	\smat{n} +   \smat{K}_{\rm om}' \smat{H}_{+}^{-1} 	\smat{n} \smat{H}_{+}^{-1} \smat{K}_{\rm om}'. 
	  \label{eq:neff}
\end{align}
These expressions reveals a clear physical picture. In the absence of optomechanical coupling i.e., $\smat{K}_\mrm{om} \!=\! 0$, the perturbation $\theta\, \smat{n}$ would act directly and locally within the system. However, the presence of the coupling matrix $\smat{K}_{\rm om}$ qualitatively changes this behavior: the perturbation is now “dressed” by virtual excursions into the complementary $(\omega\Id + \smat{H})$ sector. As a result, the system does not only respond to the bare perturbation, but also to an induced, nonlocal contribution mediated by the resolvent $\smat{H}_{+}^{-1}$. 

Since the Green's function now explicitly depends on the perturbation parameter $\theta$, all output quantities derived from it inherit the same dependence. In particular, the output covariance matrix  given in \eqnref{eq:VFout} becomes $\theta$-dependent and are accordingly denoted by $\smat{V}_\mrm{F,\theta}^\mrm{out}$ (dropping the $\omega$ dependence) as
\begin{align}
	\smat{V}_\mrm{F,\theta}^\mrm{out} &\!=\! 	(\Id-\kappa\, \smat{G}_{\theta})\, \smat{V}_\mrm{F_+}^\mrm{in}\,
	(\Id-\kappa\, \smat{G}_{\theta})^{\tp} \nonumber \\& \quad \!	+\! 	\kappa\,  \smat{G}_{\theta}\, 	\Xi\, \smat{V}_\mrm{L_{\pm},F_{-}}^\mrm{in} 
	\Xi^{\tp}	\smat{G}_{\theta}^{\tp}. \label{eq:VFoutTheta}
\end{align}
 Thus, the external perturbation influences the internal dynamical response encoded in $\smat{G}_{\theta}$, and propagates directly into the measurable output observables.

\subsection{Gaussian estimation and the role of Green's function}

In light of \eqnref{eq:Svanish}, the probe field exiting the system is fully specified by \eqnref{eq:VFoutTheta}. The parameter $\theta$ may then be estimated within Gaussian estimation theory, with the Fisher information quantifying the estimation precision.

\subsubsection{Fisher Information and precision bounds in Gaussian systems}
	Consider a family of quantum states $\rho(\theta)$ depending smoothly on an unknown parameter $\theta$. A measurement performed on the system produces an outcome $\xi$ distributed according to the conditional probability density $p(\lambda|\theta)$. The amount of information about $\theta$ that can be extracted from the measurement outcomes is quantified by the classical Fisher information (CFI)~\cite{Paris2009},
	\begin{equation}
		F(\theta)
		=
		\int d\lambda \, p(\lambda|\theta)
		\left[
		\frac{\partial}{\partial \theta}
		\ln p(\lambda|\theta)
		\right]^2 .
	\end{equation}
	The CFI determines the precision attainable for a given measurement strategy.  Optimising over all physically allowed measurements leads to the quantum Fisher information (QFI), which depends solely on $\rho_\theta$ and is defined as
	\begin{equation}
		\mathcal{F}(\theta)
		=
		\mathrm{Tr}\!\left[\rho(\theta)L_\theta^2\right],
	\end{equation}
	where $L_\theta$ denotes the symmetric logarithmic derivative (SLD), implicitly determined by
	\begin{equation}
		\frac{\partial \rho(\theta)}{\partial \theta}
		=
		\frac{1}{2}
		\Bigl(
		\rho(\theta)L_\theta
		+
		L_\theta\rho(\theta)
		\Bigr).
	\end{equation}
	
Suppose that $\nu$ independent copies of the state are measured and used to construct an unbiased estimator $\tilde{\theta}$ of the true parameter value. Its mean-squared error,
\begin{equation}
	\mathrm{Var}(\tilde{\theta}) \!=\! 	\mathbb{E}_{p(\lambda|\theta)}\!\left[(\tilde{\theta}-\theta)^2\right],
\end{equation}
where the expectation value is defined as $\mathbb{E}_{p(\lambda|\theta)}[\bullet] \!:=\! \int d\lambda\, p(\lambda|\theta)\,\bullet$. In the quantum case, this classical average is naturally replaced by the trace over the quantum state, $	\mathbb{E}_{p(\lambda|\theta)}[\bullet]
\;\longrightarrow\; \mathrm{Tr}\!\left[\rho_\theta\,\bullet\right]$.  In this setting, the variance is constrained by the hierarchy
\begin{equation}
	\mathrm{Var}(\tilde{\theta})
	\geq
	\frac{1}{\nu F(\theta)}
	\geq
	\frac{1}{\nu \mathcal{F}(\theta)}.
\end{equation}
	The first inequality is the classical Cramér--Rao bound associated with a fixed measurement strategy, whereas the second follows from the fact that no measurement can provide more information than allowed by the QFI.  It is therefore convenient to introduce the classical and quantum precision benchmarks
	\begin{equation}
		\delta_{\mathrm{C}}^2
		:=
		\frac{1}{F(\theta)},
		\qquad
		\delta_{\mathrm{Q}}^2
		:=
		\frac{1}{\mathcal{F}(\theta)},
	\end{equation}
	so that every unbiased estimation protocol obeys
	\begin{equation}
		\nu\,\mathrm{Var}(\tilde{\theta})
		\geq
		\delta_{\mathrm{C}}^2
		\geq
		\delta_{\mathrm{Q}}^2 .
	\end{equation}
	The quantity $\delta_{\mathrm{Q}}^2$ represents the ultimate sensitivity limit achievable in principle, while $\delta_{\mathrm{C}}^2$ characterises the best precision attainable with a specific measurement scheme.

For Gaussian states, the QFI can be expressed entirely in terms of the first and second moments of the quadrature operators. Let $\smat{S}$ and $\smat{V}$ denote the mean quadrature vector and covariance matrix, respectively. The QFI corresponding to parameter $\theta$ is then given by~\cite{JZhang2014}
\begin{align}\label{eq:GaussianQFI_rewritten}
	\mathcal{F}_{\theta}
	=
	\big(\partial_{\theta}\smat{S}\big)^\tp
	\smat{V}^{-1}
	\big(\partial_{\theta}\smat{S}\big)
	+
	\frac{1}{2}
	\mathrm{Tr}
	\!\left[
	\left(
	\smat{V}^{-1}
	\partial_{\theta}\smat{V}
	\right)^2
	\right].
\end{align}
The above expression naturally separates the contributions originating from the displacement of the Gaussian state in phase space and from the changes in its noise properties encoded in the covariance matrix.

For a specific measurement protocol producing Gaussian-distributed outcomes, the corresponding CFI  also assumes an analogous structure. In particular, for measurement statistics of the form
\begin{align}
	p(\bm{x}|\theta)
	&=
	\frac{
		1
	}{
		\sqrt{
			(2\pi)^n
			\det \smat{C}(\theta)
		}
	}\nonumber \\& \quad \times 
	\exp\!\left[
	-\frac{1}{2}
	\left(
	\bm{x}-\bar{\bm{x}}(\theta)
	\right)^\tp
	\smat{C}(\theta)^{-1}
	\left(
	\bm{x}-\bar{\bm{x}}(\theta)
	\right)
	\right],
\end{align}
where $\bar{\bm{x}}(\theta)$ and $\smat{C}(\theta)$ denote the mean vector and covariance matrix of the measured observables, respectively, the CFI is given by~\cite{KayBook}
\begin{align}\label{eq:GaussianCFI}
	F_{\theta}
	=
	\big(\partial_{\theta}\bar{\bm{x}}\big)^\tp
	\smat{C}^{-1}
	\big(\partial_{\theta}\bar{\bm{x}}\big)
	+
	\frac{1}{2}
	\mathrm{Tr}
	\! \left[
	\left(
	\smat{C}^{-1}
	\partial_{\theta}\smat{C}
	\right)^2
	\right].
\end{align}
As in the quantum case, the two terms separately account for sensitivity in the mean and in the covariance of the measurement outcomes.

\subsubsection{Reduction to fluctuation and dependence on Green's function}

In the present formulation, the fundamental operators are decomposed into coherent amplitudes and fluctuation operators as introduced in \eqnref{eq:alphabeta}, and all physically relevant observables are expressed in terms of the latter, implying that the mean quadrature vector vanishes identically (cf.~Eq.~\eqref{eq:Svanish}). Consequently, the first-moment contribution in Eq.~\eqref{eq:GaussianQFI_rewritten} does not contribute, and both QFI and CFI are fully determined by second-order correlations. In particular, in terms of  $\smat{V}_\mrm{F,\theta}^\mrm{out}$ from \eqnref{eq:VFoutTheta}, the QFI reduces to~\footnote{
	Although $\hat{F}_{b}$ is an auxiliary mode, one may project onto the measurable optical mode $\hat{F}_{a}$ and compute the Fisher information using only the corresponding reduced covariance matrix. Since the scaling is governed entirely by the underlying response function itself [see Eqs.~\eqref{eq:QFIpropG} and \eqref{eq:CFIpropG}], excluding $\hat{F}_{b}$ does not modify the scaling behavior of the Fisher information---the primary focus of this work---and affects only the associated prefactors.
}
\begin{equation}\label{eq:GQFIsys}
	\mathcal{F}_{\theta} = 	\frac{1}{2} \mathrm{Tr}\!\left[ \left( 	\left(\smat{V}_\mrm{F,\theta}^\mrm{out}\right)^{-1}
	\partial_{\theta}\smat{V}_\mrm{F,\theta}^\mrm{out} 	\right)^2 \right].
\end{equation}
Further insight into the scaling of $\mathcal{F}_{\theta}$ follows from the structure of the response function in Eq.~\eqref{eq:Gtheta}. In particular, the output covariance inherits its leading dependence from the Green's function, such that $\smat{V}_\mrm{F,\theta}^\mrm{out} \sim \smat{G}_\theta^2$, while its parameter derivative scales as $\partial_{\theta}\smat{V}_\mrm{F,\theta}^\mrm{out} \sim \smat{G}_\theta\,\partial_{\theta}\smat{G}_\theta$. Using the identity $\partial_\theta \smat{G}_\theta
\!=\!
-\kappa \, \smat{G}_\theta \, \smat{n}_\mrm{eff}  \, \smat{\Omega} \, \smat{G}_\theta$, 
it follows that
\begin{equation}
	\big(\smat{V}_\mrm{F,\theta}^\mrm{out}\big)^{-1}
	\partial_{\theta}\smat{V}_\mrm{F,\theta}^\mrm{out}
	\sim \smat{G}_\theta,
\end{equation}
and hence from \eqnref{eq:GQFIsys}, we note that
\begin{equation}\label{eq:QFIpropG}
	\mathcal{F}_{\theta} \sim \mathrm{Tr}\!\left[\smat{G}_\theta^2\right],
\end{equation}
demonstrating that the achievable sensitivity is governed by the square of system Green's function.

The classical measurement strategy considered in this work is the heterodyne detection of the probe output field, in which  both quadratures are measured simultaneously, resulting in an effective addition of vacuum noise (see~\secref{sec:experiment} for details). The corresponding measured amplitude and covariance matrix entering Eq.~\eqref{eq:GaussianCFI} is \cite{Weedbrook2012}
\begin{equation}\label{eq:HetDetection}
\smat{x}_\mrm{F,\theta}^\mrm{out} \!=\! \smat{S}_\mrm{F,\theta}^\mrm{out}, \qquad 	\smat{C}^\mrm{out}_\mrm{F,\theta} \!=\! \smat{V}_\mrm{F,\theta}^\mrm{out} + \Id.
\end{equation}
Owing to the same reasoning of vanishing first moments, we write
\begin{equation}\label{eq:GCFIsys}
	F_{\theta} \!=\! \frac{1}{2} \mathrm{Tr}\!\left[ \left( \left(\smat{C}_\mrm{F,\theta}^\mrm{out}\right)^{-1} \partial_{\theta}\smat{C}_\mrm{F,\theta}^\mrm{out}
	\right)^2 \right].
\end{equation}
Applying the same Green’s-function scaling arguments to the CFI, one obtains
\begin{equation}\label{eq:CFIpropG}
	F_{\theta} \sim \mathrm{Tr}\!\left[\smat{G}_\theta^2\right].
\end{equation}

In summary, both classical and quantum precision bounds are entirely determined by the output covariance matrix, whose structure is governed by the system response function $\smat{G}_\theta$. This response function therefore controls how the parameter $\theta$ is imprinted onto the measured output.

\subsection{Role of singular effective dynamical generator}\label{sec:RoleOfSingularity}

To make the role of the Green's function explicit, we first consider the regime in which the effective dynamical generator $\smat{H}_\mrm{eff}$ appearing in the Green's function in  ~\eqnref{eq:Gtheta} is invertible, i.e.
\begin{equation}
	\det \smat{H}_\mrm{eff} \neq 0.
\end{equation}
In this case, the inverse entering the Green's function admits the Neumann expansion
\begin{equation}
	\left( 	\smat{H}_\mrm{eff} 	+ 	\theta\,\smat{n}_\mrm{eff} 	\right)^{-1} 	= 	\smat{H}_\mrm{eff}^{-1} \sum_{k=0}^{\infty}
	\left( 	-\theta\, 	\smat{H}_\mrm{eff}^{-1} \,	\smat{n}_\mrm{eff} 	\right)^k.
\end{equation}
Taking the limit $\theta \to 0$ term by term, all contributions with $k \ge 1$ vanish, and only the leading term survives, yielding
\begin{equation}
	\lim_{\theta \to 0} \left( 	\smat{H}_\mrm{eff} 	+ 	\theta\,\smat{n}_\mrm{eff} 	\right)^{-1} 	= 	\smat{H}_\mrm{eff}^{-1}.
\end{equation}
Since the Green's function $\smat{G}_\theta$ is constructed from this inverse and depends on it continuously, it follows that
\begin{equation}
	\lim_{\theta \to 0} \smat{G}_\theta = -\,\smat{\Omega}\,\smat{H}_\mrm{eff}^{-1},
\end{equation}
Consequently, all output observables derived from $\smat{G}_\theta$ become independent of $\theta$ in this limit. In particular, the output amplitude and covariance matrix reduce to their unperturbed forms, implying that no residual sensitivity to $\theta$ remains at the level of the measured statistics. As a result, the QFI associated with estimation of $\theta$ is constant in this regime, reflecting the absence of any observable dependence on the perturbation parameter in the $\theta \to 0$ limit.

We now turn to the complementary regime in which the effective dynamical generator becomes singular,
\begin{equation}\label{eq:SingCond}
	\det \smat{H}_\mrm{eff} = 0.
\end{equation}
In contrast to the regular case discussed above, the inverse entering the Green's function is no longer well-defined at $\theta=0$, and the perturbation plays a crucial role in regularizing the resolvent. In this situation, the inverse admits the so called Sain--Massey expansion~\cite{Sain1969,FeiZhou},
\begin{equation}\label{eq:SMexpansion}
	\left( \smat{H}_\mrm{eff} + \theta\,\smat{n}_\mrm{eff} \right)^{-1}
	\!=
	\theta^{-s}
	\left(
	\smat{X}_0
	+
	\theta\,\smat{X}_1
	+
	\cdots
	\right),
	\quad \smat{X}_0 \!\ne\! 0.
\end{equation}
The exponent $s$ characterizes the leading divergence of the inverse and is determined by the degeneracy structure of $\smat{H}_\mrm{eff}$ at $\theta = 0$, i.e., by the dimensionality and coupling structure of its null space. Physically, it quantifies how strongly the resolvent amplifies perturbations near the singular point, with larger $s$ corresponding to stronger sensitivity to the regularizing parameter $\theta$.  The matrices $\smat{X}_0, \smat{X}_1, \ldots$ describe successive corrections to this dominant divergent contribution. They are fully determined by the structure of $\smat{H}_\mrm{eff}$ and $\smat{n}_\mrm{eff}$ and can, in principle, be obtained systematically from the resolvent expansion. In this sense, they act as generalized residues that encode both the leading divergence and the subleading response of the Green’s function. Importantly, no further assumptions beyond the existence of the expansion are required: once the singular structure is fixed, both the pole order and all coefficients are uniquely determined by the underlying dynamical generator. For detailed construction of Sain-Massey expansion, see Appendix of~\cite{Naikoo2023}.

 This non-analytic dependence on $\theta$ is consequently transferred to the Green's function, leading to a strongly enhanced sensitivity to the perturbation.  As a consequence, the QFI exhibits a divergent scaling for small $\theta$~\cite{Naikoo2023,Naikoo2025},
\begin{equation}\label{eq:QFIsing}
	\mathcal{F}_{\theta}
	\approx
	\theta^{-s}
	\left(
	A
	+
	\order{\theta}
	\right),\quad A \ne 0,
\end{equation}
where the coefficient $A$ is given by 
\begin{align}\label{eq:defA}
	A \!:= \! \Tr \left[ \Bigl(
	\smat{\Omega} (\smat{X}_{0}^{\tp})^{-1} \smat{Y}_{0}^{-1}\,
	\smat{n}\,\smat{X}_{0}\,\smat{Y}_{0}\,\smat{X}_{0}^{\tp}\,\smat{\Omega}
	\Bigr) 	+ 	\Bigl( 	\smat{\Omega}\,\smat{X}_{0}\,\smat{n}\,\smat{\Omega} \Bigr)^{\tp}\right]^2,
\end{align}
with $\smat{Y}_{0} =  (\smat{H}_\mrm{eff} \, \smat{\Omega}  +  \kappa \, \Id)  \,  \smat{V}_\mrm{F_+}^\mrm{in} \, (\smat{H}_\mrm{eff} \, \smat{\Omega}  +  \kappa \, \Id)^\tp + \kappa \, \Xi \, \smat{V}_\mrm{L_{\pm},F_{-}}  \Xi^\tp$, where $\smat{V}_\mrm{F_+}^\mrm{in}$, $\smat{V}_\mrm{L_{\pm},F_{-}}$, and $\Xi$ appear  in~\eqnref{eq:VFout}. Thus, in sharp contrast to the non-singular case—where the response becomes insensitive to $\theta$ in the limit $\theta\to0$—the singular regime leads to a divergent QFI, reflecting a strongly enhanced sensitivity of the system to the perturbation.

\section{Optomechanically Controlled Singularity and Sensing Advantage}\label{sec:optoSensing}
Having established that a divergent scaling of the QFI arises only when the effective dynamical generator $\smat{H}_\mrm{eff}$ is singular, we now demonstrate how this condition can be physically realized and controlled in an optomechanical setting. In particular, we show that the optomechanical interaction provides a tunable mechanism to access the singular regime, thereby enabling enhanced sensing performance.

The explicit form of the dynamical matrix $\smat{H}$ entering the definition of $\smat{H}_\mrm{eff}$ in \eqnref{eq:Heff} is given by 
\begin{equation}
	\smat{H}=
	\begin{pmatrix}
		\dzero - 2 \gzero  \Re(\beta) &     - \gzero |\alpha|  & \gamma_{a}       & 0 \\
		-\gzero |\alpha|       &         \omega_\mrm{M}  & 0         & \gamma_{b} \\
		-\gamma_{a}     &     0 & \dzero - 2 \gzero  \Re(\beta) & -\gzero |\alpha| \\
		0              &    -\gamma_{b} & - \gzero |\alpha|        & \omega_\mrm{M}
	\end{pmatrix}, \label{eq:Heffmatrix}
\end{equation}
where $\dzero$, $\gzero$, and $\gamma_{a(b)}$ were introduced in Eqs.~\eqref{eq:a}--\eqref{eq:b}, and $|\alpha|$ denotes the absolute value of $\alpha$ from  \eqnref{eq:alphabeta}.  
As discussed in \secref{sec:RoleOfSingularity}, the emergence of a divergent response is associated with the singularity of the effective generator. For convenience, we restate the singularity condition introduced in~\eqnref{eq:SingCond} of \secref{sec:RoleOfSingularity} in terms of the structure of $\smat{H}_\mrm{eff}$ as
\begin{equation}\label{eq:detHnew}
	\det\!\left( \smat{H}_{-} + \smat{K}_\mrm{om}^{\prime} \smat{H}_{+}^{-1}\smat{K}_\mrm{om}^{\prime} \right)=0,
\end{equation}
where $\smat{K}_\mrm{om}^{\prime} = \Omega \smat{K}_\mrm{om}$, with $\smat{K}_\mrm{om}$ defined in~\eqnref{eq:HKsFourierSpace}, encodes the optomechanical coupling between the photon and phonon sectors, and $\smat{H}_{\pm}$ are defined in~\eqnref{eq:Hpm}. This expression makes explicit that the condition involves both the uncoupled sectors and their coupling through the optomechanical interaction, whose role will be further clarified below.

To quantify the sensing performance, we consider the estimation of a real parameter $\theta$ that perturbs both the cavity detuning $\dzero$ and the mechanical frequency $\omega_\mrm{M}$, namely
\begin{equation}
	\dzero \rightarrow \dzero + \theta, \qquad \omega_\mrm{M} \rightarrow \omega_\mrm{M} + \theta. \label{eq:Idpert}
\end{equation}
This perturbation is incorporated at the level of dynamical generator by setting $\smat{n} = \smat{I}$ in the matrix $\smat{n}_\mrm{eff}$ in \eqnref{eq:neff}, which then takes the form
\begin{equation}\label{eq:neffExample}
\widetilde{\smat{n}}_\mrm{eff}  := \smat{n}_\mrm{eff}(\smat{n} = \Id)	= 	\smat{I}  - \smat{K}_{\rm om}'  (\smat{H}_{+}^{-1})^2 \smat{K}_{\rm om}'.
\end{equation}
For the present choice of perturbation, the Green's function \eqnref{eq:Gtheta} then takes the form
\begin{align}\label{eq:GthetaIdentityPert}
	\widetilde{\smat{G}}_{\theta}
	\approx & -\Omega \left[ \smat{H}_\mrm{eff} + \theta \, \widetilde{\smat{n}}_\mrm{eff} \right]^{-1}.
\end{align}
Equation~\eqref{eq:neffExample} makes explicit that a diagonal perturbation proportional to $\smat{I}$, which would otherwise shift the cavity and mechanical frequencies identically, is transformed by the optomechanical coupling $\smat{K}_\mathrm{om}$ into a generally non-diagonal effective perturbation that influences other system parameters.
\begin{table*}[htbp]
		\centering
		\renewcommand{\arraystretch}{1.6}
		\setlength{\tabcolsep}{8pt}
	\begin{tabular}{|l|l|l|c|}
		\hline 
		\textbf{Reference} & \textbf{System Parameters  $(\gamma_a, \gamma_b, \gzero,\omega_\mrm{M})/2\pi~\mrm{(Hz)}$} & \textbf{Control Variables} $|\alpha|, \Re(\beta),  \omega~ \mrm{(Hz)}$ & \textbf{Scaling}  $\delta_{Q,C}$\\ \hline  
		Murch \textit{et al.}~\cite{Murch2008NatPhys}          							       & $6.6\times 10^{5}, 1\times 10^{3}, 6 \times 10^{6}, 4.2 \times 10^{4}$         & $1.00 \times 10^{0},  3.51 \times 10^{-1}, 3.09\times 10^{6} $   & $\theta$ \\ \hline
		Chan \textit{et al.}~\cite{Chan2011Nature}                  							    & $5\times 10^{8}, 3.9 \times 10^{4}, 9\times 10^{5}, 3.9 \times 10^{9} $        & $1.38\times 10^{7}, 2.16\times 10^{2},, 2.45\times 10^{10}$     & $\theta$  \\ \hline
		Teufel \textit{et al.}~\cite{Teufel2011Nature}            								    & $2 \times 10^{5}, 32 \times 10^{0}, 2 \times 10^{2}, 1.1 \times 10^{7} $         & $5.01\times 10^{6}, 2.75 \times 10^{4}, 6.91\times 10^{7}$       &  $\theta$ \\ \hline
		Verhagen \textit{et al.}~\cite{Verhagen2012Nature}      						    & $7.1 \times 10^{6}, 3.4 \times  10^{3}, 3.4 \times 10^{3}, 7.8 \times 10^{7}$ & $5.00\times 10^{6},  1.14 \times 10^{4}, 4.9\times 10^{8} $       & $\theta$  \\ \hline
		Thompson \textit{et al.}~\cite{Thompson2008Nature}    						    & $5 \times 10^{5}, 1.2 \times 10^{-1}, 5\times 10^{1}, 1.3 \times 10^{5}$          & $1.35\times 10^{3},  1.30 \times 10^{3}, 8.17\times 10^{5} $      & $\theta$  \\ \hline
		Kleckner \textit{et al.}~\cite{Kleckner2011OE}               							  &  $4.7\times 10^{5}, 1.3\times  10^{-2}, 2.2 \times 10^{1}, 9.7\times  10^{3}$  & $2.48\times 10^{2},  2.46 \times 10^{2}, 6.09\times 10^{4} $    & $\theta$  \\ \hline
		Groblacher \textit{et al.}~\cite{Groeblacher2009Nature} 							 &$2\times 10^{5}, 1.4\times  10^{2}, 3.9 \times 10^{0}, 9.5\times 10^{5}$         &$1.25\times 10^{5},  1.21 \times 10^{5}, 5.96\times 10^{6} $       & $\theta$  \\ \hline
		 Arcizet \textit{et al.}~\cite{Arcizet2006Nature}										  & $1\times 10^{6}, 8.1\times 10^{1}, 1.2 \times 10^{0}, 8.14 \times 10^{5}$        & $3.396\times 10^{5},  3.393 \times 10^{5}, 5.11\times 10^{6} $ & $\theta$  \\ \hline
		Cuthbertson \textit{et al.}~\cite{Cuthbertson1996RSI}		& $2.75\times 10^{2}, 2.5\times 10^{-6}, 1.2\times 10^{-3}, 1\times 10^{3}$     & $4.18\times 10^{5},  4.16 \times 10^{5}, 6.28\times 10^{3} $       & $\theta$  \\ \hline
	\end{tabular}
	\caption{\textbf{System parameters and control variables for different experimental implementations.} The system parameters are taken from the corresponding experimental works reported in the literature~\cite{Aspelmeyer2014}. The control variables are chosen such that the singularity condition in~\eqnref{eq:detHnew} is satisfied. In all cases, the detuning is fixed to $\dzero = \omega_\mathrm{M}$. The perturbation $\theta$ accounts for fluctuations in $\dzero$ and $\omega_\mathrm{M}$ as described in~\eqnref{eq:Idpert}. For the representative control-variable choices listed in the table, both the classical and quantum Cramér--Rao bounds exhibit \emph{linear} scaling with $\theta$. }
	\label{tab:exps}
\end{table*}
%
%
\subsubsection{ Sensing in non-singular regime {\rm :}  $\smat{K}_\mrm{om}  =  \smat{0}$}

We begin by considering the case in which the optomechanical interaction is switched off, i.e., $\smat{K}_\mrm{om} = \smat{0}$, corresponding to $g \!=\! \gzero |\alpha|  \!=\!0$. In this limit, the photon and phonon sectors decouple, and the effective dynamical generator loses the structure responsible for the emergence of singular behavior. The determinant of this matrix then reduces to
\begin{equation}
	\det \smat{H}_\mrm{eff} =\left[\gamma_{a}^2 + \left(\omega - \dzero \right)^2 \right] \left[\gamma_{b}^2 + \left(\omega - \omega_\mrm{M} \right)^2 \right],
\end{equation}
which is strictly positive for all real values of the probe frequency $\omega$. Consequently, the  singularity condition in \eqnref{eq:SingCond} cannot be satisfied. The effective dynamical generator therefore remains nonsingular throughout this regime.

As discussed in \secref{sec:RoleOfSingularity}, the absence of singularity has direct implications for the system response. In particular, the inverse appearing in the Green's function is now regular at $\theta=0$, and admits a well-defined expansion in powers of $\theta$. Taking the limit $\theta \to 0$, the Green's function reduces smoothly to its unperturbed form and becomes effectively independent of $\theta$ in this regime. As a result, the output observables—being fully determined by the Green's function—do not exhibit any appreciable variation with respect to the parameter $\theta$.

This lack of $\theta$-dependence directly translates into a loss of sensitivity. In particular, both the CFI and QFI no longer display any divergent behaviour, but instead saturate to constant values,
\begin{equation}
	\mathcal{F}_{\theta}^\mrm{(\smat{K}_{om} = \smat{0})} \propto \theta^0, \quad F_{\theta}^\mrm{(\smat{K}_{om} = \smat{0})} \propto \theta^0.  \label{eq:FisherNS}
\end{equation}
Accordingly, the corresponding estimation errors remain finite in the limit $\theta \to 0$, without any scaling advantage.\bigskip

\subsubsection{Sensing in singular regime {\rm :}   $\smat{K}_\mrm{om} \ne \smat{0}$} \label{sec:regular}

We  now examine the system in the presence of optomechanical interaction. In this case, the coupling between the optical and mechanical modes enables access to parameter regimes where the singularity condition in \eqnref{eq:detHnew} is satisfied. More generally, the set of parameter values fulfilling this condition appears to form a hypersurface in the high-dimensional parameter space $\{\alpha, \beta, \gamma_a, \gamma_b,\gzero, \dzero, \omega, \omega_\mrm{M}\}$. Owing to many parameters involved, this hypersurface is difficult to visualize explicitly, although specific instances can be readily identified.

While experimentally relevant parameter regimes are listed in~\tabref{tab:exps}, we first consider a simplified analytically tractable case. For concreteness, we take the system tunable parameters
\begin{equation}
	\gamma_a \!=\! \gamma_b \!=\! \gzero  \!=\! \dzero \!=\!  \omega_\mrm{M} \!=\! 1, \label{eq:pars}
\end{equation}
 and set the probe frequency $\omega \!=\! \sqrt{2}$. This choice then fixes the constants $\alpha \!=\! \sqrt{2}$  and $\beta \!=\! 1$, such that \eqnref{eq:detHnew} is satisfied. For this choice of parameters, the effective dynamical matrix in \eqnref{eq:Heff} reduces to
\begin{equation}
	\smat{H}_{\mathrm{eff}}=
	\begin{pmatrix}
		a & c & -d & e\\
		c & b & e & d\\
		d & -e & a & c\\
		-e & -d & c & b
	\end{pmatrix},
	\begin{aligned}
		a&=\frac{4}{3}+\sqrt{2},
		b=-\frac{4}{3}+\sqrt{2},\\
		c&=\frac{2\sqrt{2}}{3},
		d=\frac{\sqrt{2}}{3},
		e=\frac{2}{3}.
	\end{aligned}
\end{equation}
Thus the effective dynamical matrix becomes analytically tractable in a simplified form, which directly enables a controlled expansion of its inverse. In particular, the inverse in~\eqnref{eq:GthetaIdentityPert} admits a Sain--Massey expansion of the form~\eqnref{eq:SMexpansion} about $\theta\!=\!0$, from which one obtains 
\begin{equation}
	\widetilde{\smat{G}}_\theta \approx \theta^{-2} (\smat{X}_{0} + \theta\, \smat{X}_1 ),
\end{equation}
where the leading and subleasing coefficient matrices $\smat{X}_0$ and $\smat{X}_1$ are explicitly given by
\begin{equation}
	\smat{X}_0 =
	\begin{pmatrix}
		a_0 & b_0 & c_0 & d_0 \\
		b_0 & a_0 & d_0 & g_0 \\
		e_0 & f_0 & a_0 & b_0 \\
		f_0 & -g_0 & b_0 & a_0
	\end{pmatrix}, ~~ \smat{X}_1 =
	\begin{pmatrix}
	a_1 & b_1 & c_1 & d_1 \\
	b_1 & e_1 & d_1 & a_1 \\
	e_1 & -d_1 & a_1 & b_1 \\
	-d_1 & -a_1 & b_1 & e_1
	\end{pmatrix}, 
	\label{eq:X0X1}
\end{equation}
with the elements
\begin{align}
	a_0 &= -\frac{1}{2\sqrt{2}}, \quad
	b_0  = \frac{1}{\sqrt{2}}, \quad
	c_0  = 1 - \frac{3}{2\sqrt{2}}, \quad
	d_0 = -\frac{1}{2}, \nonumber\\
	e_0 &= \frac{1}{8+6\sqrt{2}}, \quad
	f_0  = \frac{1}{2}, \quad
	g_0  = 1 + \frac{3}{2\sqrt{2}}, \nonumber\\[4pt]
	a_1 &= \frac{1}{4}(1+\sqrt{2}), \quad
	b_1  = -\frac{1}{2}, \quad
	c_1  = \frac{1}{4}(-1+\sqrt{2}), \nonumber\\
	d_1 &= \frac{1}{2\sqrt{2}}, \quad
	e_1  = \frac{1}{4}(1-\sqrt{2}).
	\label{eq:coeffs}
\end{align}
The singular structure of $\widetilde{\smat{G}}_\theta$ leads to a qualitatively different response compared to the regular case discussed in~ \secref{sec:regular}. In particular, form  \eqnref{eq:QFIsing}, it follows that  the QFI and the corresponding root mean square error scale as
\begin{equation}\label{eq:QFIanalytics}
	\mathcal{F}_{\theta}^\mrm{(\smat{K}_{om} \ne \smat{0})} \propto \theta^{-4}, \quad \delta_{Q}\theta^{(\smat{K}_\mrm{om} \ne \smat{0})} \propto \theta^{2}.
\end{equation}
The exact proportionality constant can, in principle, be obtained from~\eqnref{eq:defA} using $\smat{n} \!=\! \smat{I}$ and the coefficient $\smat{X}_{0}$ given in~\eqnref{eq:X0X1}. For the present purpose, however, it suffices to determine the asymptotic dependence on $\theta$.
A similar scaling is obtained for the CFI associated with heterodyne detection, as defined in \eqnref{eq:GaussianCFI} along with the corresponding covariance matrix given in  \eqnref{eq:HetDetection}. One obtains
\begin{equation}\label{eq:CFIanalytics}
	F_{\theta}^\mrm{(\smat{K}_{om} \ne \smat{0})} \propto \theta^{-4}, \quad \delta_{C}\theta^{(\smat{K}_\mrm{om} \ne \smat{0})} \propto \theta^{2}.
\end{equation}

\begin{figure}[t!]
	\includegraphics[width=0.9\linewidth]{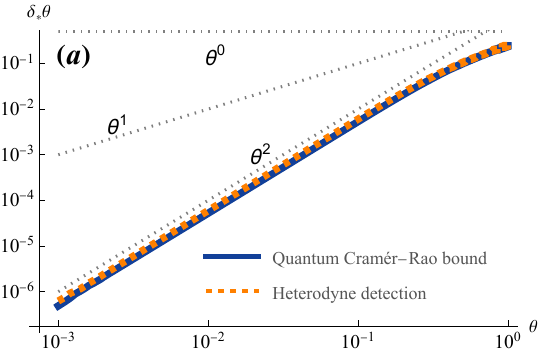}
	\includegraphics[width=0.9\linewidth]{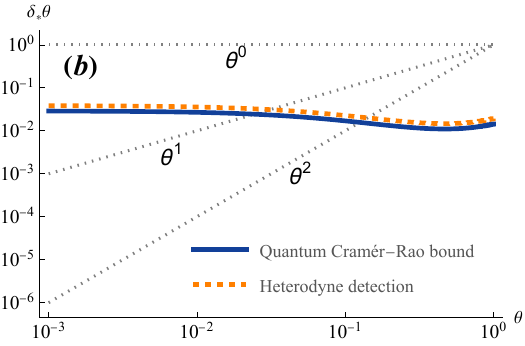}
\caption{\textbf{Root-mean-square estimation error} $\delta_{\ast}\theta$ as a function of the perturbation strength $\theta$, perturbing the detuning and mechanical frequency according to \eqnref{eq:Idpert}. The solid blue and dashed orange curves denote the quantum and classical bounds, $\delta_Q\theta$ and $\delta_C\theta$, obtained from Eqs.~\eqref{eq:GQFIsys} and \eqref{eq:GCFIsys}, respectively. 
	\textbf{(\textit{a})} $\smat{H}_\mrm{eff}$ in \eqnref{eq:Heffmatrix} is tuned to a singular configuration using system parameter given in \eqnref{eq:pars} along with the control variable $\alpha, \beta$ and probe frequency $\omega$. Both the quantum and classical bounds exhibit the \emph{quadratic} scaling $\delta_{\ast}\theta\propto\theta^2$, in agreement with the analytical predictions of Eqs.~\eqref{eq:QFIanalytics} and \eqref{eq:CFIanalytics}, thereby demonstrating consistent scaling behavior for both the quantum-limited and measurement-constrained estimation strategies. 
	\textbf{(\textit{b})} Same parameters as in panel \textbf{(\textit{a})}, except for $\alpha \!\approx \!0$, rendering $\smat{H}_\mrm{eff}$ nonsingular. In this case, the estimation error becomes insensitive to $\theta$ in the limit $\theta\to0$, consistent with analytical predictions in \eqnref{eq:FisherNS}. 
	For both panels, the input fields are taken to be thermal  states with covariance matrices $\smat{V}_\mrm{F}^\mrm{in}[\pm\omega]\!=\!2\,n_{\pm\omega}+\Id$ for the probe modes and $\smat{V}_\mrm{L}^\mrm{in}[\pm \omega]\!=\!2\,m_{\pm \omega}+\Id$ for the bath mode. The occupation numbers are chosen as $n_{\omega}\!=\!m_{\omega}\!=\!1$, and $n_{-\omega}\!=\! m_{- \omega} \!=\!0$. The grey dotted lines indicate the reference scalings $\theta^{0}$, $\theta^{1}$, and $\theta^{2}$.}
	\label{fig:error}
\end{figure}

In contrast to the nonsingular case, optomechanical coupling enables access to a regime in which the system response becomes highly sensitive to perturbations. This enhanced sensitivity thus arises as a direct consequence of the optomechanically induced singular structure. The comparison highlights the central role of the coupling strength: while the absence of optomechanical coupling precludes the existence of any parameter choice that reaches the singular regime, increasing the coupling makes such singular operating points accessible, where enhanced sensitivity and divergent precision can be achieved.

The above analytical predictions are further supported by numerical results in ~\figref{fig:error}, obtained directly from \eqnref{eq:GQFIsys} and its classical (heterodyne) counterpart in \eqnref{eq:GCFIsys} with covariance matrix given by \eqnref{eq:HetDetection}. In particular, the numerical evaluation of both the CFI and QFI clearly reproduces the expected scaling behaviour. When the optomechanical coupling is present and the system is tuned to the singular point, the corresponding root mean square errors exhibit the predicted \emph{quadratic} scaling with $\theta$, thereby corroborating the analytical results. In contrast, upon switching off the coupling, the numerical results display a clear saturation, consistent with the absence of any $\theta$-dependence in the regular regime.

To place these results in an experimentally relevant context, we refer to~\tabref{tab:exps}, which compiles representative optomechanical implementations across different platforms spanning several orders of magnitude in system parameters, together with control parameters chosen to satisfy the singularity condition. In all cases shown, the root mean square error exhibits a linear scaling in $\theta$, in contrast to the quadratic scaling obtained in the simplified analytical model. We emphasize, however, that the choice of control variables in~\tabref{tab:exps} satisfying the singularity condition is not unique; alternative parameter choices may also exist and could in principle lead to different scaling behavior, although identifying such solutions typically requires extensive numerical exploration.

These findings indicate that the enhanced sensitivity in the generic system considered in this work is intrinsically tied to the optomechanically induced singularity of the effective dynamical generator. Taken together, our analysis demonstrates that optomechanical interaction constitutes a key resource for engineering singular response, enabling substantial improvements in sensing performance. 

\subsection{Experimental relevance of heterodyne detection in optomechanical systems}\label{sec:experiment}

In cavity optomechanical experiments, the mechanical displacement is ultimately inferred from the phase and amplitude fluctuations imprinted on the outgoing optical field. A widely used detection strategy for accessing both quadratures simultaneously is the heterodyne detection.

\begin{figure}[t!]
	\includegraphics[width=0.75\linewidth]{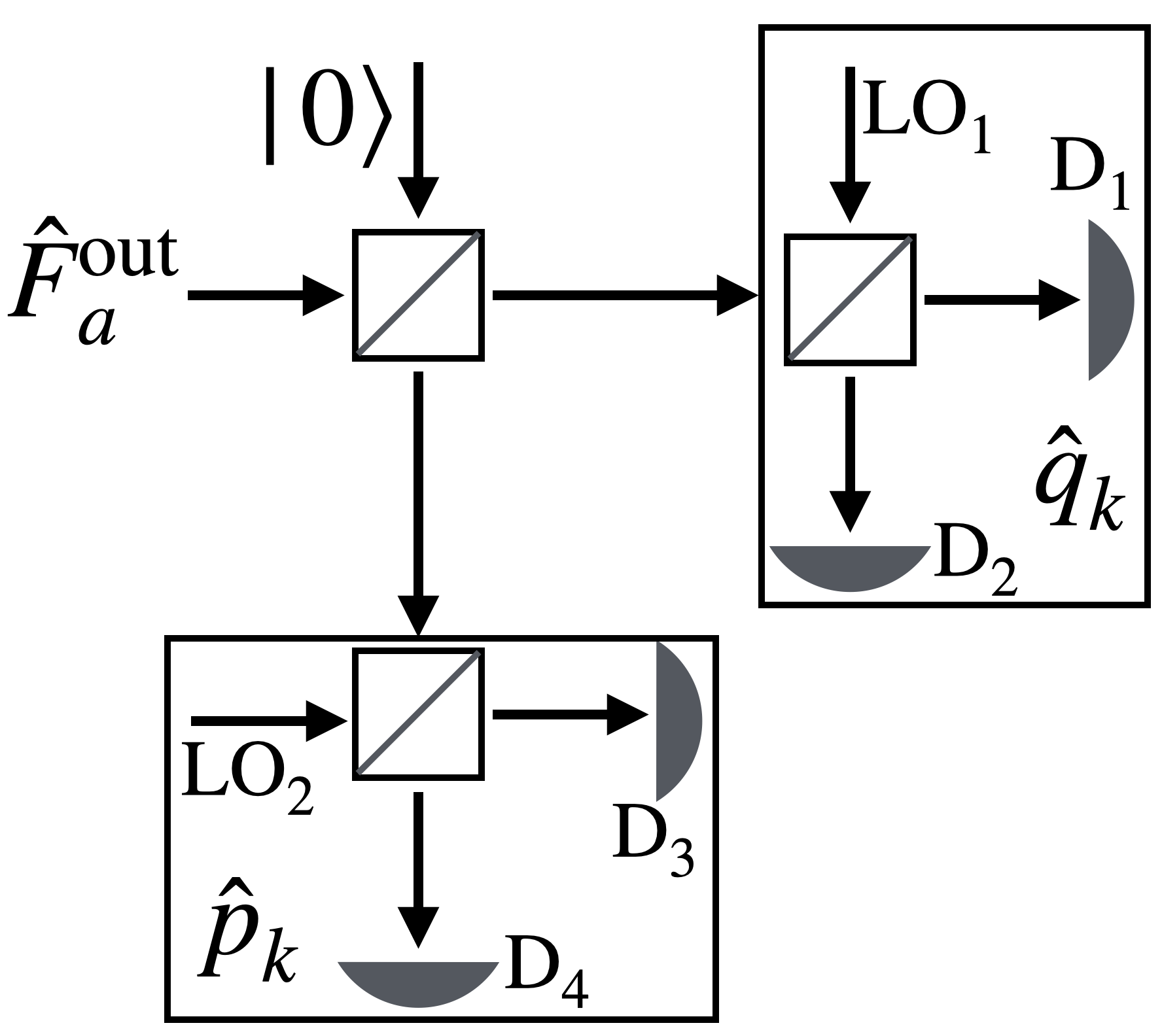}
	\caption{\textbf{Schematic of heterodyne detection via dual balanced homodyne measurements.}
		The probe output mode is first combined with vacuum at a beam splitter, generating two output ports. Each port is subsequently measured independently using balanced homodyne detection. The first output is combined with a local oscillator $\mathrm{LO}_1$ on a second beam splitter, and the resulting outputs are measured by detectors $D_1$ and $D_2$, which access the quadrature $q_k$. The second output is combined with a local oscillator $\mathrm{LO}_2$ on another beam splitter, with detection performed by $D_3$ and $D_4$, which measure the quadrature $p_k$. Together, these two homodyne measurements realize a heterodyne detection scheme that reconstructs both conjugate quadratures of the probe field.}
	\label{fig:heterodyne}
\end{figure}

In a typical implementation, illustrated schematically in~\figref{fig:heterodyne}, the output field from the optomechanical cavity is first combined with vacuum at a balanced 50:50 beam splitter, generating two output ports that are subsequently analyzed through independent balanced homodyne measurements~\cite{Walton2021}. Each output is mixed with a strong local oscillator on a second beam splitter and detected via a pair of photodiodes, thereby enabling separate measurements of the two conjugate field quadratures. Together, the dual homodyne measurements realize a heterodyne detection scheme that simultaneously reconstructs both quadratures of the optical field. Consequently, heterodyne detection provides access to the full Gaussian statistics of the output field, at the expense of an additional unit of vacuum noise [cf.~\eqnref{eq:HetDetection}] arising from the vacuum mode introduced at the initial beam splitter~\cite{Ferraro2005,Weedbrook2012}.

In optomechanical systems, heterodyne detection and related dual-homodyne measurement schemes have become standard tools for probing mechanical motion and extracting the full Gaussian statistics of the output field. Such techniques have played a central role in experiments on ground-state cooling and sideband thermometry~\cite{Teufel2011,JBClark2017}, including recent demonstrations of high-purity room-temperature quantum optomechanics~\cite{Dania2025}. Heterodyne measurements have furthermore been extensively employed for the observation and interpretation of motional sideband asymmetry and quantum noise signatures in cavity optomechanical systems~\cite{Borkje2016}, as well as for high-precision displacement sensing and the recovery of quantum correlations in optomechanical spectra~\cite{Monteiro2017}. These measurement protocols are now routinely implemented across a broad range of optomechanical platforms, including membrane-in-the-middle and photonic-crystal cavity systems~\cite{Aspelmeyer2014,SafaviNaeini2013}.

\section{Conclusion}\label{sec:concl}
In this work, we developed a theoretical framework for analyzing parameter estimation in optomechanical systems based on a frequency-domain Green's function formalism combined with Gaussian quantum estimation theory. Within this framework, perturbations acting  on the system degrees of freedom are encoded in the dynamical response and become accessible through measurements on the output cavity field. This approach provides a direct connection between the response properties of the driven optomechanical system and the achievable precision in parameter estimation.

Our analysis shows that the optomechanical interaction fundamentally modifies the structure of the system Green's function and enables operation in the vicinity of an effectively singular configuration. Perturbations that shift the system toward or away from this configuration produce disproportionately large variations in the response function and, consequently, in the output field statistics. As a result, the information carried by the output field about the perturbation parameter is strongly amplified. This amplification manifests itself in a divergent scaling of  the QFI and the corresponding estimation precision. Furthermore, the CFI associated with heterodyne detection exhibits the same asymptotic scaling as the quantum counterpart, demonstrating that this experimentally accessible measurement scheme captures the full scaling advantage provided by the amplified response.

Away from the regime of strong optomechanical coupling, the response no longer exhibits the near-singular behavior responsible for the enhanced sensitivity. In particular, for vanishing optomechanical coupling the Green's function remains regular and analytic in the perturbation parameter, leading to a saturation of both the QFI and CFI.

These findings demonstrate that optomechanical interactions provide a controllable mechanism for accessing regimes of enhanced dynamical susceptibility, in which weak perturbations induce amplified changes in the output field and consequently yield improved parameter-estimation precision.
	
\section*{Acknowledgments} 
 This work was supported by the Czech Science Foundation under Project No.~25--15775S. The author gratefully acknowledges Jan Pe\v{r}ina for his valuable comments and  suggestions.

	%

	\appendix
	\twocolumngrid
	\setcounter{secnumdepth}{3}
	
\section{Linearised optomechanical dynamics}\label{app:model}

We consider a standard cavity optomechanical system consisting of a single optical mode coupled to a mechanical oscillator. Throughout, we set $\hbar = 1$. The system Hamiltonian in the laboratory frame reads~\cite{Aspelmeyer2014}
\begin{align}
	\hat{H} &= \omega_\mrm{cav}\, \hat{a}^\dagger \hat{a} + \omega_\mrm{M}\, \hat{b}^\dagger \hat{b}
	- \gzero\, \hat{a}^\dagger \hat{a}(\hat{b}^\dagger + \hat{b}) \nonumber \\
	&\quad + \i \, \varepsilon_\mrm{L}\, \left(e^{-\i \omega_\mrm{L} t}\hat{a}^\dagger - e^{\i \omega_\mrm{L} t}\hat{a}\right).  \label{eq:Hlab}
\end{align}
Here, $\hat{a}$ ($\hat{a}^\dagger$) and $\hat{b}$ ($\hat{b}^\dagger$) denote the annihilation (creation) operators of the cavity and mechanical modes with resonance frequencies $\omega_\mrm{cav}$ and $\omega_\mrm{M}$, respectively. The third term describes the radiation--pressure interaction between the optical and mechanical degrees of freedom, with $g_0$ being the vacuum optomechanical coupling strength. The last term accounts for the coherent driving of the cavity mode by an external laser field of frequency $\omega_\mrm{L}$ and driving amplitude $\varepsilon_\mrm{L}$.  To eliminate the explicit time dependence, we move to a frame rotating at the laser frequency using the unitary transformation $\hat{U} = e^{\i \omega_\mrm{L} t \hat{a}^\dagger \hat{a}}$. The transformed Hamiltonian becomes
\begin{align}
	\hat{H}' &= \hat{U}\hat{H}\hat{U}^\dagger + \i\, (\partial_t \hat{U})\hat{U}^\dagger \nonumber \\
	&= -\dzero \, \hat{a}^\dagger \hat{a} + \omega_\mrm{M} \, \hat{b}^\dagger \hat{b}
	- \gzero\, \hat{a}^\dagger \hat{a}(\hat{b}^\dagger + \hat{b}) \nonumber \\
	&\quad + \i \,\varepsilon_\mrm{L} \,(\hat{a}^\dagger - \hat{a}),
\end{align}
where $\dzero = \omega_\mrm{L} - \omega_\mrm{cav}$ denotes the laser--cavity detuning.

To account for dissipation, we couple both the optical and mechanical modes to independent Markovian reservoirs within the standard input--output framework. For each mode, we distinguish between two physically distinct types of loss channels. The rates $\kappa_a$ and $\kappa_b$ describe the coupling of the cavity mode $\hat{a}$ and the mechanical mode $\hat{b}$ to accessible probe field modes, with corresponding input operators $\hat{F}_a^{\mathrm{in}}$ and $\hat{F}_b^{\mathrm{in}}$. These channels represent ports through which the system can be measured. In addition, the rates $\eta_a$ and $\eta_b$ account for intrinsic losses arising from uncontrolled interactions with other environmental degrees of freedom, described by independent noise operators $\hat{L}_a^{\mathrm{in}}$ and $\hat{L}_b^{\mathrm{in}}$. This decomposition allows us to explicitly separate engineered coupling to external fields from uncontrollable dissipation.

Defining the total damping rates $\gamma_\ell = (\kappa_\ell + \eta_\ell)/2$ for $\ell = a,b$, we introduce the combined input noise operators
	\begin{equation}
		\hat{\xi}_a^{\mathrm{in}} = \sqrt{\kappa_a}\hat{F}_a^{\mathrm{in}} + \sqrt{\eta_a}\hat{L}_a^{\mathrm{in}}, 
		\quad
		\hat{\xi}_b^{\mathrm{in}} = \sqrt{\kappa_b}\hat{F}_b^{\mathrm{in}} + \sqrt{\eta_b}\hat{L}_b^{\mathrm{in}},
	\end{equation}
	which collect the contributions from the probe and intrinsic loss channels. The Heisenberg--Langevin equations then read
	\begin{align}
		\partial_t \hat{a} &= (-\i\, \dzero - \gamma_a)\hat{a}
		+ \i\,  \gzero\, \hat{a}(\hat{b}^\dagger + \hat{b}) + \varepsilon_\mrm{L} + \hat{\xi}_a^{\mathrm{in}}, \\
		\partial_t \hat{b} &= (-\i \, \omega_\mrm{M} - \gamma_b)\hat{b}
		+ \i \, \gzero\,  \hat{a}^\dagger \hat{a} + \hat{\xi}_b^{\mathrm{in}}.
	\end{align}

To proceed further, we decompose the operators into classical steady-state amplitudes and quantum fluctuations,
\begin{align}
	\hat{a}(t) = \alpha \,  \mathbb{1} + \delta\hat{a}(t), \qquad
	\hat{b}(t) = \beta\,  \mathbb{1} + \delta\hat{b}(t),
\end{align}
where $\mathbb{1}$ is the identity operator, $\alpha$ and $\beta$ are complex numbers, and $\delta\hat{a}$, $\delta\hat{b}$ describe fluctuations about the steady state. Substituting this decomposition into the equations of motion and collecting terms of equal order yields the mean-field (c-number) equations
\begin{align}
	\dot{\alpha} &= (-\i \, \dzero - \gamma_a)\alpha
	+ \i \, \gzero \alpha(\beta + \beta^*) + \varepsilon_\mrm{L}, \label{eq:alpha_app2}\\
	\dot{\beta} &= (-\i \, \omega_\mrm{M} - \gamma_b)\beta + \i \, \gzero \, |\alpha|^2. \label{eq:beta_app2}
\end{align}
In steady state, $\dot{\alpha} = \dot{\beta} = 0$. In the experimentally relevant regime where the single-photon coupling is very small, $\gzero \ll 1$, the nonlinear term in Eq.~\eqref{eq:alpha_app2} can be neglected to leading order. This yields the approximate solution
\begin{align}
	\alpha \approx \frac{\varepsilon_\mrm{L}}{-\i \, \dzero - \gamma_a},
\end{align}
which is obtained from Eq.~\eqref{eq:alpha_app2} by neglecting the $\gzero$-dependent term, i.e., assuming $\gzero \ll 1$.  From this expression, the intracavity photon number is
\begin{align}
	|\alpha|^2 \approx \frac{|\varepsilon_\mrm{L}|^2}{\dzero^2 + \gamma_a^2}. \label{eq:alphasq}
\end{align}
In typical experiments, strong driving implies $|\alpha|^2 \gg 1$, often reaching values on the order of $10^8$, corresponding to a highly populated cavity mode. Substituting \eqnref{eq:alphasq} into \eqnref{eq:beta_app2}, we obtain
\begin{align}
	\beta \approx \frac{\i \, \gzero}{\i\, \omega_\mrm{M} + \gamma_b}
	\frac{|\varepsilon_\mrm{L}|^2}{\dzero^2 + \gamma_a^2}.
\end{align}
This shows that the mechanical oscillator acquires a steady-state displacement proportional to the intracavity photon number. Physically, this corresponds to radiation pressure exerted by the cavity field shifting the equilibrium position of the mechanical mode.

\begin{table*}[ht!]
	\centering
	\small
	\renewcommand{\arraystretch}{1.25}
	\begin{tabular}{lll}
		\hline
		Regime & Description & Physical implications \\
		\hline
		
		Linearized regime &
		\shortstack[l]{\rule{0pt}{2.6ex}Driven linear optomechanics: fluctuations around steady state,\\
			valid even for small $\bar{n}_{\rm cav}$ if photons are unresolved ($\kappa$ large)} &
		\shortstack[l]{\rule{0pt}{2.6ex}Linear response approximation\\holds beyond large-photon limit} \\
		\hline
		
		Strong coupling ($g>\gamma_{a}$) &
		\shortstack[l]{\rule{0pt}{2.6ex}Normal-mode hybridization via coherent\\light–mechanics interaction} &
		\shortstack[l]{\rule{0pt}{2.6ex}Coherent exchange between cavity\\field and mechanical oscillator} \\
		\hline
		
		Single-photon regime ($\gzero>\gamma_{a}$) &
		\shortstack[l]{\rule{0pt}{2.6ex}Intrinsic nonlinear interaction where individual\\photons significantly modify mechanics} &
		\shortstack[l]{\rule{0pt}{2.6ex}Onset of observable quantum\\nonlinear optomechanics} \\
		\hline
		
		Red detuning ($\dzero \approx -\omega_\mrm{M}$) &
		\shortstack[l]{\rule{0pt}{2.6ex}Beam-splitter interaction\\
			$- g(\delta \hat{a}^\dagger \, \hat{b} + \delta \hat{a} \, \hat{b}^\dagger)$\\
			enabling resonant excitation exchange} &
		\shortstack[l]{\rule{0pt}{2.6ex}Sideband cooling and state transfer\\(RWA, $\gamma_{a} \ll \omega_\mrm{M}$)} \\
		\hline
		
		Blue detuning ($\dzero \approx +\omega_\mrm{M}$) &
		\shortstack[l]{\rule{0pt}{2.6ex}Two-mode squeezing\\
			$- g(\delta \hat{a}^\dagger \,  \hat{b}^\dagger + \delta \hat{a} \, \hat{b})$\\
			generating correlated photon–phonon pairs} &
		\shortstack[l]{\rule{0pt}{2.6ex}Parametric amplification, entanglement,\\and dynamical instability} \\
		\hline
		
		Zero detuning ($\dzero = 0$) &
		\shortstack[l]{\rule{0pt}{2.6ex}Radiation-pressure coupling\\
			$- g(\delta \hat{a}^\dagger+\delta \hat{a})(\hat{b}+\hat{b}^\dagger)$\\
			linking mechanical displacement to optical phase} &
		\shortstack[l]{\rule{0pt}{2.6ex}Displacement sensing and QND\\measurement of optical amplitude quadrature} \\
		\hline
		
	\end{tabular}
	\caption{\textbf{Interaction regimes} in the cavity optomechanics framework~\cite{Aspelmeyer2014}.}
	\label{tab:regimes}
\end{table*}

We now turn to the fluctuation dynamics given by
\begin{align}
	\partial_t \delta\hat{a} &= \Big[(-\i \, \dzero - \gamma_a)
	+ \i \, \gzero(\beta + \beta^*)\Big] \, \delta\hat{a} \nonumber \\
	&\quad + \i \, \gzero \alpha (\delta\hat{b} + \delta\hat{b}^\dagger)
	+ \i \,\gzero \,  \delta\hat{a}(\delta\hat{b} + \delta\hat{b}^\dagger) + \hat{\xi}_a^\mrm{in}, \\
	\partial_t \delta\hat{b} &= (-\i \, \omega_\mrm{M} - \gamma_b) \, \delta\hat{b}
	+ \i \,\gzero  \, (\alpha^*\delta\hat{a} + \alpha \delta\hat{a}^\dagger) \nonumber \\
	&\quad + \i \,\gzero \,  \delta\hat{a}^\dagger \delta\hat{a}
	+ \hat{\xi}_b^\mrm{in}.
\end{align}
The above equations contain nonlinear fluctuation terms such as $\delta\hat{a}^\dagger \delta\hat{a}$ and $\delta\hat{a}(\delta\hat{b} + \delta\hat{b}^\dagger)$. In the regime of large intracavity amplitude $|\alpha| \gg 1$, these terms are small compared to the linear terms proportional to $\alpha$ and can be safely neglected and this constitutes the \emph{linearisation} approximation. Table~\ref{tab:regimes} summarizes the main interaction regimes of cavity optomechanics and their characteristic physical phenomena. 

Retaining only linear terms, we obtain
\begin{align}
	\partial_t \delta\hat{a} &= \Big[(-\i\,\dzero - \gamma_a)
	+ \i \, \gzero(\beta + \beta^*)\Big] \, \delta\hat{a} \nonumber \\
	&\quad + \i \, \gzero  \, \alpha (\delta\hat{b} + \delta\hat{b}^\dagger)
	+ \hat{\xi}_a^\mrm{in}, \\
	\partial_t \delta\hat{b} &= (-\i \, \omega_\mrm{M} - \gamma_b)\, \delta\hat{b}  \nonumber \\
	&\quad 
	+ \i \, \gzero \, (\alpha^*\delta\hat{a} + \alpha \delta\hat{a}^\dagger)+ \hat{\xi}_b^\mrm{in}.
\end{align}
It is convenient to remove the phase of $\alpha$ by writing $\alpha = |\alpha| e^{\i\varphi}$ and defining the rotated operator $\hat{\tilde{a}} = e^{-\i\varphi}\delta\hat{a}$. This transformation renders the coupling real and equations then take the form
\begin{align}
	\partial_t \hat{\tilde{a}}  &= (-\i \, \Delta  - \gamma_a) \hat{\tilde{a}} 
	+ \i \, g \, (\hat{b} + \hat{b}^\dagger) +  \hat{\tilde{\xi}}_a^\mrm{in}, \\
	\partial_t \hat{b} &= (-\i \, \omega_\mrm{M}  - \gamma_b)\hat{b}
	+ \i \, g \, (\hat{\tilde{a}}  + \hat{\tilde{a}} ^\dagger) + \hat{\xi}_{b}^\mrm{in},
\end{align}
where we have introduced the \emph{effective} detuning and \emph{enhanced} coupling strength,
\begin{align}
	\Delta &= \dzero - 2\, \gzero \, \mathrm{Re}(\beta), \\
	g &= \gzero \, |\alpha|.
\end{align}
The effective detuning includes a radiation-pressure-induced frequency shift, while the coupling $g$ is enhanced by the large intracavity amplitude.  Finally, dropping the tilde notation for simplicity, the linearised equations can be written as
\begin{align}\label{ap:adotbdot}
	\partial_t \hat{a} &= (-\i \, \Delta - \gamma_a)\,\hat{a}
	+ \i \, g \, (\hat{b} + \hat{b}^\dagger)  + \hat{\xi}_{a}^\mrm{in}, \\
	\partial_t \hat{b} &= (-\i \, \omega_\mrm{M} - \gamma_b)\, \hat{b}
	+ \i \, g \, (\hat{a} + \hat{a}^\dagger) + \hat{\xi}_{b}^\mrm{in}.
\end{align}
These equations form the starting point for the analysis developed in the main text. One can immediately see that the averages are given by
\begin{align}
	\partial_t \avg{\hat{a}} &= (-\i \, \Delta - \gamma_a)\avg{\hat{a}}
	+ \i\,  g\, \avg{(\hat{b} + \hat{b}^\dagger)}, \\
	\partial_t \avg{\hat{b}} &= (-\i \, \omega_\mrm{M} - \gamma_b)\avg{\hat{b}}
	+ \i \, g \, \avg{(\hat{a} + \hat{a}^\dagger)}.
\end{align}
These equations can be derived from the effective quadratic Hamiltonian
\begin{align}
	\hat{\tilde{H}} &= - \Delta\,  \hat{a}^\dagger \hat{a}
	+ \omega_\mrm{M} \, \hat{b}^\dagger \hat{b} + g \, (\hat{a} + \hat{a}^\dagger)(\hat{b} + \hat{b}^\dagger),
\end{align}
which provides a convenient starting point for analyzing quantum fluctuations, correlations, and noise properties of the optomechanical system.

\section{Frequency-domain formulation and output covariance matrix}\label{app:QuadRep}
In this section, we reformulate the dynamics in the frequency domain and derive the output covariance matrix underlying the metrological analysis of the main text. For the Gaussian fluctuation states considered here, the first moments vanish, and the relevant information is completely encoded in the covariance matrix. The frequency-domain formulation provides a convenient route to its evaluation from the linearized Heisenberg--Langevin equations. To this end, we introduce the Fourier-domain representation of the bosonic operator vector $\bvec{a}(t) = (\hat{a}(t), \hat{b}(t), \ldots)^\tp$ as
\begin{equation}
	\bvec{a}[\omega] := \mathcal{F}_\omega[\bvec{a}] = 	\int dt \, e^{\i \omega t} \, \bvec{a}(t).
\end{equation}
This convention implies that Hermitian conjugation acts in frequency space as
\begin{equation}
	\big(\bvec{a}[\omega]\big)^\dagger = 	\bvec{a}^\dagger[-\omega],
\end{equation}
where the dagger is understood to act entry-wise on the vector components. 

Starting from the time-domain Heisenberg equations of motion for the annihilation operators in~Eq.~\eqref{eq:veca} of the main text, we obtain the corresponding linear algebraic relations in frequency space,
\begin{align}
	-\i \,  \omega \, \bvec{a}[\omega] &= -\i\,\H \, \bvec{a}[\omega] +\K_\mrm{om} \bvec{a}^{\dagger}[\omega] \nonumber \\ 
	& \quad +\K_\mrm{probe} \bvec{F}^\mrm{in}[\omega] +\K_\mrm{bath} \bvec{L}^\mrm{in}[\omega],\label{apeq:veca} 
\end{align}
where we have used that $\partial_t \to -\i \, \omega$ under the Fourier transform convention adopted above. We further define the frequency-domain quadrature operators as
\begin{align}
	\q[\omega] &:= \bvec{a}[\omega] + \big(\bvec{a}[\omega]\big)^\dagger, \nonumber \\
	\p[\omega] &:= -\,\i \, \big(\bvec{a}[\omega] - \big(\bvec{a}[\omega]\big)^\dagger \big).
\end{align}
With these definitions, we obtain from Eqs.~\eqref{apeq:veca}, and its conjugate, the following relations:
\begin{align}
	\omega\, \p[\omega]  &= \Re(\H) \p[\omega] + \Im(\H) \q[\omega] +  \Re(\K_\mrm{om}^*) \q[-\omega] \nonumber \\& \quad - \Im(\K_\mrm{om}^*) \p[-\omega] + \K_\mrm{probe} \q_\mrm{F}^\mrm{in}[\omega]  + \K_\mrm{bath} \q_\mrm{L}^\mrm{in}[\omega] \label{apeq:pomega}\\
	- \omega\, \q[\omega]	&= -\Re(\H) \, \q[\omega] + \Im(\H) \, \p[\omega]  -  \Re(\K_\mrm{om}^*) \, \p[-\omega] \nonumber \\& \quad - \Im(\K_\mrm{om}^*) \, \q[-\omega]   + \K_\mrm{probe} \, \p_\mrm{F}^\mrm{in}[\omega] + \K_\mrm{bath} \, \p_\mrm{L}^\mrm{in}[\omega] \label{apeq:qomega}
\end{align}

In the notation introduced in Eq.~\eqref{eq:FSnotation} of the main text, i.e.
\begin{equation}\label{apeq:FSnotation}
	\qsvec S[\pm \omega] \!:=\!
	\begin{pmatrix}
		\q[\pm \omega]\vspace{1mm} \\
		\p[\pm \omega]
	\end{pmatrix}, \quad 
	\qsvec{S}^\mrm{in/ out}_{\ast}[ \omega] \!:=\! \begin{pmatrix}
		\q_{\ast}^\mrm{in/out}[\omega] \vspace{1mm} \\
		\p_{\ast}^\mrm{in/out}[\omega]
	\end{pmatrix},
\end{equation}
where $\ast\!=\! \mrm{F/L}$ pertain to the probe field and inaccessible channels, respectively, Eqs.~\eqref{apeq:pomega} and \eqref{apeq:qomega} can be compactly written as
\begin{align}
	\omega\,\Omega\,\qsvec{S}[\omega]
	&=
	\smat{H}\,\Omega\,\qsvec{S}[\omega]
	+
	\smat{K}_\mrm{om}\,\qsvec{S}[-\omega]
	\nonumber\\
	&\quad+
	\smat{K}_\mrm{probe}\,\qsvec{S}_\mrm{F, in}[\omega]
	+
	\smat{K}_\mrm{bath}\,\qsvec{S}_\mrm{L, in}[\omega],
	\label{apeq:1_re3}
\end{align}
Here, $\smat{H}$ denotes the dynamical matrix governing the coherent evolution of the quadratures, while $\smat{K}_\mrm{probe}$ and $\smat{K}_\mrm{bath}$ characterize the coupling of the intracavity modes to the probe and bath degrees of freedom, respectively. The matrix $\smat{K}_\mrm{om}$ originates from the optomechanical interaction and explicitly couples the positive- and negative-frequency sectors.

Since the equations of motion are not closed within the positive-frequency sector alone, one must also consider the corresponding equation for the negative-frequency quadratures. Proceeding as above, one obtains
\begin{align}
	-\omega\,\Omega\,\qsvec{S}[-\omega]
	&=
	\smat{H}\,\Omega\,\qsvec{S}[-\omega]
	+
	\smat{K}_\mrm{om}\,\qsvec{S}[\omega]
	\nonumber\\
	&\quad+
	\smat{K}_\mrm{probe}\,\qsvec{S}_\mrm{F, in}[-\omega]
	+
	\smat{K}_\mrm{bath}\,\qsvec{S}_\mrm{L, in}[-\omega].
	\label{apeq:2_re3}
\end{align}
Equations~\eqref{apeq:1_re3} and \eqref{apeq:2_re3} therefore constitute a coupled linear system in the frequency domain. Eliminating the negative-frequency quadrature $\qsvec{S}[-\omega]$ from Eq.~\eqref{apeq:1_re3} using Eq.~\eqref{apeq:2_re3}, one arrives at an equation involving only the positive-frequency intracavity quadratures,
\begin{align}
	\qsvec{S}[\omega]
	&=
	\smat{G}[\omega]\,\smat{K}_\mrm{probe}\,\qsvec{S}_\mrm{F, in}[\omega]
	+
	\smat{G}[\omega]\,\smat{K}_\mrm{bath}\,\qsvec{S}_\mrm{L, in}[\omega]
	\nonumber\\
	&\quad
	-
	\smat{G}[\omega]\,\smat{T}_\mrm{op}\,\qsvec{S}_\mrm{F, in}[-\omega]
	-
	\smat{G}[\omega]\,\smat{T}_\mrm{ob}\,\qsvec{S}_\mrm{L, in}[-\omega].
	\label{apeq:Sw2_re3}
\end{align}
This  corresponds to \eqnref{eq:Sw2} of the main text with matrices  $\smat{T}_\mrm{op}$, $\smat{T}_\mrm{ob}$ and  $\smat{G}[\omega]$  given in Eqs.~\eqref{eq:Top}, \eqref{eq:Tob}, and \eqref{eq:Gfun}, respectively.

The observable quantities of interest are the output probe fields, which are related to the intracavity modes through the standard input-output formalism~\cite{GardinerBook}. For the probe channels associated with the cavity mode $\hat a$ and  mechanical mode $\hat b$, the input-output relations are
\begin{equation}
	\hat{F}_{a}^\mrm{out} 	= 	\hat{F}_{a}^\mrm{in} - 	\sqrt{\kappa_a}\,\hat a, \qquad \hat{F}_{b}^\mrm{out} = \hat{F}_{b}^\mrm{ in}
	- 	\sqrt{\kappa_b}\,\hat b.
\end{equation}
 In the quadrature representation introduced in \eqnref{eq:FSnotation}, the above equations can be written compactly in Fourier space as
\begin{equation}
	\qsvec{S}_\mrm{F}^\mrm{out}[\omega] = \qsvec{S}_\mrm{F}^\mrm{in}[\omega] - 	\smat{K}_\mrm{probe}\,\qsvec{S}[\omega].
	\label{apeq:IOrel_re3}
\end{equation}
Substituting the intracavity solution \eqnref{apeq:Sw2_re3} into \eqnref{apeq:IOrel_re3}, and assuming $\kappa\!:=\! \kappa_a\!=\! \kappa_b$, yields the output probe quadratures entirely in terms of the input noise operators,
\begin{align}
	\qsvec{S}_\mrm{F}^\mrm{out}[\omega] &= 	\left(\Id-\kappa\, \smat{G}[\omega]\right) \qsvec{S}_\mrm{F}^{\rm in}[\omega] 	-
	\sqrt{\kappa}\,\smat{G}[\omega]\,\smat{K}_\mrm{bath}\,
	\qsvec{S}_\mrm{L}^{\rm in}[\omega]
	\nonumber\\
	&\quad 	- \sqrt{\kappa}\,\smat{G}[\omega]\,\smat{T}_\mrm{op}\, \qsvec{S}_\mrm{F}^\mrm{in}[-\omega] 	- 	\sqrt{\kappa}\,\smat{G}[\omega]\,\smat{T}_\mrm{ob}\,
	\qsvec{S}_\mrm{L}^\mrm{in}[-\omega].
	\label{apeq:SFout_re3}
\end{align}
Thus, the outgoing probe fields are influenced not only by the positive-frequency components of the input fluctuations, but also by the negative-frequency sector. These arise from the parametric terms in Eq.~\eqref{eq:veca}, which couple $\bvec{a}$ and $\bvec{a}^\dagger$ and thereby mix creation- and annihilation-type fluctuations.

The linearity of the equations of motion ensures that the transformation between input and output quadratures can be represented by a scattering matrix, thereby enabling a direct construction of the output statistical moments. To this end, it is convenient to collect all input and output quadratures into a single vector and introduce the global scattering relation
\begin{equation}
	\begin{pmatrix}
		\qsvec{S}_\mrm{F}^\mrm{out}[\omega]\\
		\qsvec{S}_\mrm{L}^\mrm{out}[\omega]\\
		\qsvec{S}_\mrm{F}^\mrm{out}[-\omega]\\
		\qsvec{S}_\mrm{L}^\mrm{out}[-\omega]
	\end{pmatrix}
	=
	\mathcal{S}[\omega]
	\begin{pmatrix}
		\qsvec{S}_\mrm{F}^\mrm{in}[\omega]\\
		\qsvec{S}_\mrm{L}^\mrm{in}[\omega]\\
		\qsvec{S}_\mrm{F}^\mrm{in}[-\omega]\\
		\qsvec{S}_\mrm{L}^\mrm{in}[-\omega]
	\end{pmatrix}.
\end{equation}
The scattering matrix $\mathcal{S}[\omega]$ therefore characterizes the complete linear transformation between the incoming and outgoing quadrature operators. Its upper  blocks, relevant for our calculation are given by
\begin{equation}
	\mathcal{S}[\omega] = 	\left( 	\begin{array}{c|c}
		\Id-\kappa\, \smat{G}[\omega]
		&
		-\sqrt{\kappa}\,\smat{G}[\omega]\, \Xi
		\\
		\hline
		\dots & \dots
	\end{array}
	\right),
	\label{eq:Smatrix_re3}
\end{equation}
where
\begin{equation}
	\Xi_{4\times 12}
	:=
	\left(
	\smat{K}_\mrm{bath}
	~|~
	\smat{T}_\mrm{op}
	~|~
\smat{T}_\mrm{ob}
	\right).
\end{equation}

The covariance matrix of the output fields follows directly from the linearity of the scattering relation. The covariance matrix transforms according to
\begin{equation}
	\smat{V}^\mrm{out}_\mrm{Total}[\omega] 	= 	\mathcal{S}[\omega]\, \smat{V}^\mrm{in}_\mrm{Total}[\omega]\,
	\mathcal{S}[\omega]^\mrm{T},
\end{equation}
which is the standard transformation rule for Gaussian states under linear bosonic operations. The total input covariance matrix takes a block-diagonal form,
\begin{equation}
	\smat{V}^\mrm{in}_\mrm{Total}[\omega]
	=
	\begin{pmatrix}
		\smat{V}_\mrm{F_{+}}^\mrm{in} & 0\\
		0 & \smat{V}_\mrm{L_{\pm},F_{-}}^\mrm{in}
	\end{pmatrix},
\end{equation}
consistent with the independent preparation of the probe and bath inputs.  The  matrix $\smat{V}_\mrm{F_{+}}^\mrm{in}\!:=\! \smat{V}_\mrm{F}^\mrm{in}[\omega]$ and the lower block is given by
\vspace{0.25mm}
\begin{equation}
	\smat{V}_\mrm{L_{\pm},F_{-}}^\mrm{in}
	:=
	\smat{V}_\mrm{L}^\mrm{in}[\omega]
	\oplus
	\smat{V}_\mrm{F}^\mrm{in}[-\omega]
	\oplus
	\smat{V}_\mrm{L}^\mrm{in}[-\omega].
\end{equation}
 Using the block structure of $\mathcal{S}[\omega]$, one obtains the covariance matrix of the outgoing probe fields as 
 \vspace{0.25mm}
\begin{align}
	\smat{V}_\mrm{F}^\mrm{out}[\omega]
	&=
	(\Id-\kappa\, \smat{G}[\omega])
	\,\smat{V}_\mrm{F_+}^\mrm{in}\,
	(\Id-\kappa\, \smat{G}[\omega])^\mrm{T}
	\nonumber\\
	&\quad+
	\kappa\,
	\smat{G}[\omega]\,
	\Xi\,
	\smat{V}_\mrm{L_{\pm},F_{-}}^\mrm{in}
	\Xi^\mrm{T}\,
	\smat{G}[\omega]^\mrm{T}.
	\label{eq:Vout_re3}
\end{align}
This expression corresponds to \eqnref{eq:VFout} of the main text. The first term describes the transformation of the incoming probe fluctuations through the cavity response, while the second term accounts for the additional noise contribution originating from the bath and the interaction-induced coupling to the negative-frequency sector.  The probe and bath inputs are assumed to be thermal  states characterized by  covariance matrices 
\vspace{0.25mm}
\begin{equation}
	\smat{V}_\mrm{F}^\mrm{in}[\pm\omega]
	=
	(2n_{\pm\omega}+1)\,\Id,
	\quad
	\smat{V}_\mrm{L}^\mrm{in}[\pm\omega]
	=
	(2m_{\pm\omega}+1)\,\Id,
\end{equation}
with occupation numbers  $n_{\pm\omega}$ and $m_{\pm\omega}$, respectively. Finally, since the system is probed exclusively with positive-frequency fields, the negative-frequency input probe is taken to be in the vacuum state, implying
\vspace{0.25mm}
\begin{equation}
	n_{-\omega}=0.
\end{equation}
Under this assumption, all nontrivial correlations in the output fields arise from the positive-frequency probe fluctuations together with the frequency-mixing processes induced by the optomechanical interaction.
\end{document}